\documentclass[12pt]{article}
\usepackage{latexsym}
\usepackage{amsmath,amssymb,amsfonts,amsthm}
\usepackage[english]{babel}
\newcommand{\be}{\begin{equation}}
\newcommand{\ee}{\end{equation}}
\newcommand{\beq}{\begin{equation}}
\newcommand{\eeq}{\end{equation}}
\newcommand{\bea}{\begin{eqnarray}}
\newcommand{\eea}{\end{eqnarray}}
\newcommand{\nn}{\nonumber \\}
\newcommand{\p}[1]{(\ref{#1})}
\newcommand{\lb}{\label}

\begin{document}
\begin{titlepage}
\rightline{JINR-E2-2012-89}

\vfill

\begin{center}
\baselineskip=16pt {\Large\bf
Super Landau Models on Odd Cosets}
\vskip 0.3cm {\large {\sl }} \vskip 10.mm {\bf M.
Goykhman$^{*\,a}$, $\;$  E. Ivanov$^{\dagger\,b}$,  $\;$ S. Sidorov$^{\dagger\,c}$
}
\vspace{1cm}

${}^*${\it
Institute Lorentz for Theoretical Physics, Leiden University,\\
 P.O. Box 9506, Leiden 2300RA, The Netherlands\\
\vspace{0.2cm}

${}^\dagger$ Bogoliubov Laboratory of Theoretical Physics, JINR, \\
141980 Dubna, Moscow Region, Russia\\
}
\end{center}
\vfill

\par
\begin{center}
{\bf ABSTRACT}
\end{center}
\begin{quote}

We construct $d{=}1$ sigma models of the Wess-Zumino type on the
$SU(n|1)/U(n)$ fermionic cosets. Such  models can be regarded as a
particular supersymmetric extension (with a target space
supersymmetry) of the classical Landau model, when a charged
particle possesses only fermionic coordinates. We consider both
classical and quantum models, and prove the unitarity of the quantum
model by introducing the metric operator on the Hilbert space of the
quantum states, such that all their norms become positive-definite. It is remarkable that the quantum $n{=}2$ model
exhibits hidden $SU(2|2)$ symmetry. We also discuss the planar limit
of these models. The Hilbert space in the planar $n{=}2$ case is shown to carry $SU(2|2)$
symmetry which is different from
that of the $SU(2|1)/U(1)$ model.

\vfill \vfill \vfill \vfill \vfill \hrule width 5.cm \vskip 2.mm
{\small
\noindent $^a$ goykhman89@gmail.com\\
\noindent $^b$ eivanov@theor.jinr.ru\\
\noindent $^c$ sidorovstepan88@gmail.com\\
}
\end{quote}
\end{titlepage}

\setcounter{footnote}{0}

\setcounter{page}{1}

\numberwithin{equation}{section}

\section{Introduction}
The renowned Landau model \cite{Landau1930} describes a charged
non-relativistic particle moving on a two-dimensional Euclidean
plane $\mathbb{R}^2 \sim (z, \bar z)$ under the influence of a uniform
magnetic field which is orthogonal to the plane. Its simplest
generalization to a curved manifolds is the Haldane model
\cite{Haldane} describing a charged particle on the
two-sphere $S^2 \sim SU(2)/U(1)$ in the field of Dirac monopole
located at the center. These models are exactly solvable, on both the
classical and the quantum levels. They can be interpreted as
one-dimensional nonlinear sigma models with the Wess-Zumino (WZ) terms.
For instance, the Haldane model is described by the $d{=}1$ $SU(2)/U(1)$ sigma model action with the $U(1)$ WZ term.

There are two different approaches to supersymmetrizing this bosonic system. One is based on a worldline supersymmetry
(see, e.g., \cite{Witten}), while the other deals with a target-space supersymmetry. The latter option corresponds to extending
the bosonic manifolds to supermanifolds by adding extra fermionic target coordinates. These supermanifolds are identified
with cosets of some supergroups,
so the relevant invariant actions describe $d{=}1$ WZ sigma models on supergroups. Since supergroups possess a wider set of cosets
as compared to their bosonic subgroups, there are several non-equivalent super Landau models associated with the same supergroup.
A minimal superextension of $SU(2)$ is the supergroup $SU(2|1)$ involving the bosonic subgroup $U(2) = SU(2)\times U(1)$
and a doublet of fermionic generators.\footnote{This is a minimal possibility if one assumes
the standard complex conjugation for the generators.} It possesses a few different cosets, each giving rise to some
super Landau $d{=}1$ sigma model. The $SU(2)/U(1)$ model can be promoted either to a model on the $(2|2)$ dimensional
supersphere $SU(2|1)/U(1|1)$ \cite{CPn1,BCIMT} or to a model on the $(2|4)$ dimensional superflag manifold
$SU(2|1)/[U(1)\times U(1)]$ \cite{IMT2,BCIMT}.
Both models are exactly solvable, like their bosonic prototypes, and exhibit further interesting properties \cite{BCIMT}.
For instance, for some special relations between the coefficients of the corresponding WZ terms they prove
to be quantum-mechanically equivalent to each other.
Another surprising feature of these models is that, contrary to the standard lore, the presence of non-canonical fermionic
terms of the second-order in time derivative in their actions does not necessarily lead, upon quantization, to ghosts states
with negative norms;
all norms can be made positive-definite by modifying the inner product on the Hilbert space.

The supergroup $SU(2|1)$ also possesses the pure odd supercoset $SU(2|1)/U(2)$ of the dimension $(0|4)$.
The Landau-type quantum sigma models on the odd cosets $SU(n|1)/U(n)$ of the dimension $(0|2n)$, with the pure WZ term as the action,
were studied in \cite{Odd} (see also \cite{CPn1}). The relevant Hilbert spaces studied there
involve only single (vacuum) states associated with the lowest Landau levels (LLL).
These LLL states reveal interesting $SU(n|1)$ representation content.

There remained a problem of finding out the complete sigma model actions on the supercoset $SU(n|1)/U(n)$, such that they contain both
the WZ term and the term bilinear in the coset Cartan forms (i.e. the one-dimensional pullback of the Killing form on $SU(n|1)/U(n)$),
and of exploring the relevant quantum mechanics. The basic aim of the present paper is to fill this gap.

In section 2, we construct the $SU(n|1)$ invariant action for this model, using the $d{=}1$ version of the universal gauge approach which
can be traced back to the construction of the two-dimensional bosonic $\mathbb{CP}^n$ sigma model actions in \cite{CPn}.
The same general approach nicely works
in some other cases elaborated as instructive examples in the Appendix (Landau-type models on the bosonic coset
$SU(n+1)/U(n) \sim \mathbb{CP}^n$ and
the supersphere $SU(2|1)/U(1|1) \sim \mathbb{CP}^{(1|1)}\,$).
In section 3 we construct the corresponding Hamiltonian and $SU(n|1)$ Noether charges, in both the classical and the quantum cases.
In section 4 we present the complete set of eigenfunctions of the Hamiltonian and compute its spectrum.
The salient feature of the quantum case is that the number of LL states is finite and equal to $n+1$.
We discuss the $SU(2|1)$ representation assignment of the full set of states in the $n{=}2$ case, by computing,
in particular, the eigenvalues of the $SU(2|1)$ Casimir operators. In section 5 we compute the
norms of the LL states and find that for some values of the WZ term strength $\kappa$ there
are states with the negative and/or zero norms, like in other super Landau models. In section 6 we give a more detailed treatment
of the $n{=}2$ case. We present the explicit form of the metric operator which allows one to make all norms positive-definite.
At each LL, the quantum states are found to form short multiplets of some hidden $SU(2|2)$ symmetry which is an extension
of the original $SU(2|1)$ symmetry (the phenomenon of such an enhancement of  $SU(2|1)$ to  $SU(2|2)$ at the quantum level
was earlier revealed in the superflag Landau
model \cite{BCIMT}). In section 7 we study the planar limit of the odd coset Landau models. We show, in particular,
that the Hilbert space
for $n{=}2$ also carries some extended $SU(2|2)$ symmetry.
We finish with conclusions and outlook in section 8.

\setcounter{equation}{0}
\section{$SU(n|1)/U(n)$ action from $U(1)$ gauging}
The supergroup $SU(n|1)$ can be defined as the set of linear
transformations of the $n+1$-component multiplet $(z, \xi^i)$ ($i
= 1,2, \ldots n$), such that they preserve its norm
\be
z\bar{z}- \zeta \cdot \bar {\zeta }=inv. \label{ocinv}
\ee
Here, the components $z, \bar z$ are Grassmann-even, while $\zeta^i, \bar \zeta_i$ are
Grassmann-odd. Hereafter, $\zeta \cdot \bar {\zeta } = -\bar\zeta \cdot \zeta  =\zeta^i\bar {\zeta }_i\,$.
The fermionic transformations are:
\be
\delta_\epsilon z = \zeta\cdot\bar\epsilon\,, \quad \delta \zeta^i = \epsilon^i z\,, \lb{sun1tran}
\ee
where $\epsilon^i, \bar\epsilon_i$ are Grassmann-odd parameters. The variables $\zeta_i$ transform in the fundamental
representation
of the group $U(n)\,$. These  $U(n)$ transformations are contained in the closure of \p{sun1tran}.

Now we are going to show how, starting from this linear $SU(n|1)$ multiplet, one can construct a nonlinear $d{=}1$  WZ
sigma model action associated with the odd $(0|2n)$-dimensional coset $SU(n|1)/U(n)$.

We start with the following $SU(n|1)$ invariant Lagrangian
\be
L=\nabla z\bar{\nabla} \bar {z}+\bar {\nabla }\bar {\zeta }\cdot \nabla \zeta +2\kappa {\cal A}\,,   \label{ocilagr1}
\ee
where
\be
\nabla =\partial _t -i {\cal A}\,, \quad \bar {\nabla }=\partial _t +i {\cal A}\,.
\ee
The auxiliary gauge field ${\cal A}(t)$ ensures
the invariance of the Lagrangian \p{ocilagr1} under the $U(1)$ gauge transformations:
\be
\delta z = i\lambda z\,, \quad \delta \zeta^i = i \lambda \zeta_i\,, \quad \delta {\cal A} = \dot \lambda\,. \lb{gauge}
\ee
The last term in \p{ocilagr1} is the Fayet-Iliopoulos (FI) term, it is invariant under \p{gauge} up to a total time derivative.
The gauge field ${\cal A}(t)$ is a $SU(n|1)$ singlet; gauge transformations \p{gauge} commute with the rigid $SU(n|1)$ ones.

As the next steps, we impose the $SU(n|1)$ invariant constraint on the variables
$z, \zeta ^i$
\be
z\bar {z}- \zeta \cdot \bar{\zeta }=1\,, \label{coinvc}
\ee
choose the $U(1)$ gauge
\be
z = \bar z \equiv \rho \lb{gauge1}
\ee
and, using \p{coinvc}, express $\rho$ in terms of the fermionic variables,
\be
\rho = \sqrt {1+ \zeta\cdot\bar {\zeta}}\,.
\ee
Finally, the auxiliary field ${\cal A}$ can be eliminated by its algebraic equation of motion:
\be
{\cal A} = -\frac{1}{2}\left[2\kappa + i(\bar\zeta\cdot \dot{\zeta} - \dot{\bar\zeta}\cdot \zeta) \right].
\ee
Upon substituting all this back into the Lagrangian \p{ocilagr1}, the latter takes the form
\bea
L &=& \dot {\bar {\zeta }}\cdot \dot {\zeta } + \frac{2 + \zeta\cdot \bar\zeta}{4(1 + \zeta\cdot \bar\zeta)}
\left[(\dot\zeta\cdot\bar\zeta)^2 + (\zeta\cdot\dot{\bar\zeta})^2\right]
-\frac{\zeta\cdot\bar\zeta}{2(1 + \zeta\cdot \bar\zeta)}(\dot\zeta\cdot\bar\zeta)(\zeta\cdot\dot{\bar\zeta}) \nn
&& +\,i \kappa \left(\dot\zeta\cdot\bar\zeta - \zeta\cdot\dot{\bar\zeta} \right). \label{ocilagr2}
\eea

This Lagrangian is invariant, up to a total time derivative, under the following purely fermionic nonlinear
realization of the odd $SU(n|1)$ transformations:
\be
\delta \zeta^i = \sqrt{1 + \zeta\cdot \bar\zeta}\,\epsilon^i + \frac{\zeta^i}{2\sqrt{1 + \zeta\cdot \bar\zeta}}
\left(\bar\epsilon\cdot \zeta +  \epsilon\cdot \bar\zeta\right).
\ee
It precisely coincides with the one considered in \cite{Odd}. This realization follows from \p{sun1tran}
with taking into account the gauge \p{gauge1}, the constraint \p{coinvc} and the necessity to accompany
the original $SU(n|1)$ transformations by the compensating gauge transformations \p{gauge} with
$$
\lambda = - \frac{i}{2\sqrt{1 + \zeta\cdot \bar\zeta}}\left(\bar\epsilon\cdot \zeta +  \epsilon\cdot \bar\zeta\right)
$$
in order to preserve the gauge \p{gauge1}. The Lagrangian \p{ocilagr2} describes $d{=}1$ WZ sigma model on the odd coset $SU(n|1)/U(n)$,
with the $d{=}1$ fields $\zeta^i$ and $\bar\zeta_i$ ($2n$ real fermionic variables) being the coset parameters.
The number of independent variables was reduced from the $(n+1)$ complex ones $z, \zeta^i$ to $n$ such variables $\zeta^i$
by imposing the
constraint \p{coinvc} and choosing the gauge \p{gauge1}. In this aspect, the method we applied is quite similar
to the gauge approach to the
construction of the bosonic $d=2$ $\mathbb{CP}^n$ sigma models in \cite{CPn}. It is worth pointing out that our $d{=}1$
gauging procedure
automatically yields not only the standard sigma model part of the Lagrangian but also the WZ term with the strength
$2\kappa$.\footnote{See Appendix for further examples.} In the ``parent'' linear sigma model action \p{ocilagr1}, this constant
appears
as a strength of the FI term. In ref. \cite{Odd}, only the WZ term in \p{ocilagr2} was considered. Such truncated Lagrangian
can be treated as the large $\kappa$ limit of \p{ocilagr2}.

In what follows, it will be convenient to deal with the coset coordinates $\xi^i$ related to $\zeta^i$ by
\be
\xi ^i=\frac{\zeta ^i}{\sqrt {1+ \zeta \cdot \bar {\zeta }} }   \label{mfrepl}
\ee
and possessing the simple holomorphic transformation law
\be
\delta \xi ^i=\epsilon^i+(\bar {\epsilon }\cdot \xi )\,\xi^i. \label{suptr}
\ee
In terms of $\xi^i$, the Lagrangian \p{ocilagr2} is rewritten as:
\be
L=\frac{\dot {\bar {\xi }}\cdot \dot {\xi }}{1- \xi\cdot \bar {\xi }}+\frac{(\dot {\bar {\xi }}\cdot\xi )(\dot {\xi }\cdot\bar {\xi })}
{\left(1- \xi \cdot \bar {\xi }\right)^2}-i\kappa\,\frac{\bar {\xi }\cdot \dot {\xi }-
\dot {\bar {\xi }}\cdot \xi}{1- \xi \cdot \bar {\xi }}.   \label{Mflag}
\ee

The odd $SU(n|1)$ transformations may be expressed in terms of the fermionic $SU(n|1)$ generators $Q_i, \;  Q^{\dagger j}$:
\be
\delta \xi =\left[(\epsilon Q)+(\bar\epsilon  Q^{\dagger})\right]\xi,
\ee
where $Q_i$ and $Q^{\dagger i} = (Q_i )^\dag $ satisfy the relations:
\be
\left\{Q_i,Q_j\right\} =0, \quad \left\{Q^{\dagger i}, Q^{\dagger j}\right\}=0\, , \lb{QQ}
\ee
\be
\left\{Q_i, Q^{\dagger j}\right\}=\left(J_i^j-\delta _i^j\frac{1}{n}J_k^k \right)+\delta^j_i\,B\, . \lb{QbarQ}
\ee
Here $J_i^j-\frac{1}{n}\delta _i^jJ_k^k \equiv \tilde{J}^i_j$ are the $SU(n)$ generators, and $B$ is the generator
of the $U(1)$ transformation.
They constitute the bosonic ``body''  $U(n)$ of the supergroup $SU(n|1)$.
In the realization on the variables $\xi^i, \bar\xi_k\,$, these generators are:
\be
J_i^j=\bar \xi _i\frac{\partial}{\partial \bar \xi _j}-\xi ^j\frac{\partial}{\partial \xi ^i},
\quad
B= \left(\frac{1}{n}-1\right)J_k^k.
\ee

The explicit expressions for the conserved Noether supercharges corresponding to the odd $SU(n|1)$ transformations are
\bea
&Q_i =-\pi_i+\bar {\xi }_i \left(\bar {\xi}\cdot \bar {\pi}\right)-i\kappa\,\bar {\xi }_i , \label{ocQ}\\
&Q^{\dagger i}=\bar {\pi }{}^i+\xi^i\left(\xi \cdot\pi\right)+i\kappa\, \xi^i.  \label{ocdagQ}
\eea
Here, $\pi_i = \partial L/\partial \dot {\xi }^i$ and $\bar{\pi}{}^i = \partial L/ \partial \dot {\bar{\xi }}_i \,$
are the momenta canonically conjugate to the fields $\xi ^i$ and $\bar\xi _i$.
The corresponding conserved $U(n)$ generators ${J}^i_j$ and $B$ are expressed as
\be
{J}^j_i = \bar\xi _i\bar\pi ^j-\xi ^j\pi _i\,, \quad B
= \left(\frac{1}{n}-1\right)\left(\bar\xi _i\bar\pi ^i-\xi ^i\pi _i\right)+\text{constant} \,, \lb{Jclass}
\ee
where a constant in $B$ (the central charge)
will be fixed in the quantum model by requiring the generators to close on the $\textit{su}(2|1)$ algebra as in \p{QQ}, \p{QbarQ}.

\setcounter{equation}{0}

\section{Quantization}

\subsection{Hamiltonian formulation}

To quantize the classical $SU(n|1)/U(n)$ coset model constructed in the previous section,
we have to build its Hamiltonian formulation.
First of all, it is convenient to rewrite the Lagrangian \p{Mflag} in a geometric way, in terms of the metric on
the coset space $SU(n|1)/U(n)$ and the external gauge potentials given on this manifold. The corresponding Hamiltonian
will have the form convenient for quantization, and it will be easy to find its spectrum.

We can write down the metric on the  $SU(n|1)/U(n)$ coset parametrized by $\xi ^i$ coordinates, using the K\"ahler potential
\be
K=\ln \left(1- \xi \cdot \bar{\xi }\right)\,.
\ee
The metric is given by
\be
g^i_j =\partial _j\partial ^{\bar {i}} K =
\frac{\delta _j^i}{1- \xi \cdot\bar {\xi }}-\frac{\xi ^i\bar {\xi }_j }{\left(1- \xi \cdot\bar{{\xi}} \right)^2}\,. \lb{Metr}
\ee
Also we define the gauge connections
\be
A_i =-i\partial _i K= i\,\frac{\bar {\xi }_i }{1- \xi \cdot\bar {\xi }}\,; \quad  \bar{A}^i=i\partial ^{\bar {i}}K
=i \frac{\xi ^i}{1- \xi \cdot\bar {\xi }}\,.
\label{AhatA}
\ee
Note, that $\bar {A}^i=-\overline{({A}_i)}$. The inverse metric is given by:
\be
(g^{-1})^i_j = \left( {1- \xi \cdot\bar {\xi }} \right)\left( {\delta _j^i + \xi^i\bar {\xi }_j } \right). \lb{InvMetr}
\ee

In terms of these quantities, the Lagrangian \p{Mflag} can be written as:
\be
L=g^i_j \dot {\bar {\xi }}_i\dot {\xi }^j +\kappa (\dot {\xi }^iA_i +\dot{\bar {\xi }}_i \bar{A}^i).\label{odd-L1}
\ee

The momenta canonically conjugate to the variables $\xi^i$ and $\bar\xi_i$ are
\be
\pi _i =\frac{\partial L}{\partial \dot {\xi }^i}=-g^k_i \dot {\bar {\xi}}_k +\kappa A_i ,
\quad \bar {\pi }^i=\frac{\partial L}{\partial \dot {\bar{\xi }}_i }=g^i_k \dot {\xi }^k+\kappa \bar {A}^i.
\ee
Then the Hamiltonian is given by
\be
H=(g^{-1})^i_j (\bar {\pi }^j-\kappa \bar {A}^j)(\pi _i -\kappa A_i ). \label{occlham}
\ee

Using the anticommutativity of the Grassmann variables, we rewrite the Hamiltonian in the form
\be
H=\frac{1}{2}(g^{-1})^i_j \left[\bar {\pi}^j-\kappa \bar {A}^j ,\;\pi _i -\kappa A_i \right]
\ee
and perform canonical quantization in the coordinate representation by replacing
\beq
\pi _i \to-i\partial _i ,\;  \bar {\pi }^i\to -i\partial ^{\bar {i}}\,.
\eeq
As a result, we obtain the quantum Hamiltonian
\be
H=\frac{1}{2}(g^{-1})^i_j \left[ {\nabla ^{(\kappa )}_i ,\; \nabla^{(\kappa )\bar {j}}} \right],\lb{QH}
\ee
where we have introduced
\be
\nabla ^{(\kappa )}_i =\partial _i +\frac{\kappa \bar {\xi }_i }{1- \xi\cdot\bar {\xi }}\,, \quad
\nabla ^{(\kappa )\bar {j}}=\partial ^{\bar{j}}+\frac{\kappa \xi ^j}{1- \xi \cdot\bar {\xi }}\,.  \label{longd}
\ee
These ``semi-covariant'' derivatives satisfy the following anticommutation relations\footnote{The semi-covariant
derivatives are essentially complex, in accordance with the fact that the wave superfunctions are complex.
Under the complex conjugation they are transformed as $(\nabla ^{(\kappa )}_i, \nabla ^{(\kappa )\bar {j}})$ $\Rightarrow
\pm(\nabla ^{(-\kappa ) \bar {i}}, \nabla ^{(-\kappa )}_{j})$, when acting, respectively, on Grassmann-odd or Grassmann-even
superfunctions.}
\be
\label{ant}
\left\{ {\nabla ^{(\kappa )}_i ,\;  \nabla ^{({\kappa })}_j }\right\}=0\,, \quad
\left\{ {\nabla ^{(\kappa )\bar {i}},\;  \nabla ^{({\kappa})\bar {j}}} \right\}=0\,, \quad
\left\{ {\nabla ^{(\kappa )}_i , \nabla^{({\kappa }')\bar {j}}} \right\}=(\kappa +{\kappa }')g^j_i\,.
\ee
Using them,  we can rewrite the Hamiltonian \p{QH} in the convenient equivalent form:
\be
H=(g^{-1})^i_j \nabla ^{(\kappa )}_i \nabla ^{(\kappa )\bar {j}}-\kappa n \equiv H' - \kappa n\,. \lb{QH1}
\ee

For what follows, it will be useful to write $H'$ explicitly:
\be
H' = (1-\xi\cdot\bar\xi)\left(\partial_i\partial^{\bar i} + \xi^i \bar\xi_{\bar j}\partial_i\partial^{\bar j} \right)
+ \kappa\,(1-\xi\cdot\bar\xi)\left(\bar\xi_i\partial^{\bar i}  - \xi^j\partial_j\right) -\kappa^2\,
\xi\cdot\bar\xi + \kappa n\,. \lb{Hprime}
\ee
The Hamiltonian is hermitian with respect to the appropriate inner product (see below).

\subsection{Quantum $SU(n|1)$ generators}
The quantum $SU(n|1)$ generators can be obtained from the classical expressions \p{ocQ}, \p{ocdagQ} and \p{Jclass}
\bea
&& Q_i=-\partial _i+\bar\xi _i\bar\xi _j\partial ^{\bar j}+\kappa\bar\xi _i\,, \quad
Q^{\dagger i}=-\partial ^{\bar i}-\xi ^i\xi ^j\partial _j+\kappa\xi ^i , \lb{Qkvan} \\
&& J_i^j=\bar \xi _i\frac{\partial}{\partial \bar \xi _j}-\xi ^j\frac{\partial}{\partial \xi ^i} \equiv \tilde{J}_i^j
+ \frac{1}{n} \delta_i^j\,J_k^k\,,
\quad
F= \left(\frac{1}{n}-1\right)J_k^k -2\kappa\,, \lb{JBkvan}
\eea
where we properly fixed the ordering ambiguities, based on the same reasonings as in \cite{Odd}.
A constant in the expression for the Noether charge $B$ (see \p{Jclass}) was fixed to be $-2\kappa$
and the resulting operator was denoted $F$ in order to have
the $su(n|1)$ algebra \p{QQ}, \p{QbarQ} in the quantum case:
\be
\left\{Q_i,Q_j\right\} =0, \quad \left\{Q^{\dagger i}, Q^{\dagger j}\right\}=0\, , \lb{QQ}
\ee
\be
\left\{Q_i, Q^{\dagger j}\right\}= \tilde{J}_i^j +\delta^j_i\,F\, . \lb{QbarQ}
\ee
Using the explicit form \p{Hprime} of $H'$,
it is straightforward to check that the $SU(n|1)$
generators \p{Qkvan}, \p{JBkvan} indeed commute with the Hamiltonian.

For further use, we explicitly present the generators in the $n{=}2$ case.
The corresponding Landau model possesses the eight-parameter symmetry supergroup $SU(2|1)$.

We define
\be
\Pi =Q_1\,, \quad Q=Q_2\,.
\ee

Then the full set of the $SU(2|1)$ generators as differential operators acting on the manifold with the odd coordinates
$(\xi^i, \bar\xi_i), i=1,2\,,$,
is given by the expressions
\bea
&&Q=-\partial _2 +\bar {\xi }_2 \bar {\xi }_1 \partial ^{\bar {1}}+\kappa\bar {\xi }_2,\\
&&\Pi =-\partial _1 +\bar {\xi }_1 \bar {\xi }_2 \partial ^{\bar {2}}+\kappa\bar {\xi }_1,\\
&&J_+ =i\bar {\xi }_2 \partial ^{\bar {1}}-i\xi ^1\partial _2,\\
&&J_- =i\xi ^2\partial _1 -i\bar {\xi }_1 \partial ^{\bar {2}}, \\
&&J_3 =\frac{1}{2}(\xi ^1\partial _1 -\xi ^2\partial _2 -\bar {\xi }_1\partial ^{\bar {1}}+\bar {\xi }_2 \partial ^{\bar {2}}), \\
&&F=\frac{1}{2}(\xi ^1\partial _1 +\xi ^2\partial _2 -\bar {\xi }_1 \partial^{\bar {1}}-\bar {\xi }_2 \partial ^{\bar {2}})-2\kappa\,.
\eea
The corresponding non-vanishing (anti)commutation relations read
\be
\left[J_+,J_-\right] = 2J_3\, , \qquad \left [J_3,J_\pm\right] =
\pm J_\pm\,,\lb{JJ}
\ee
\bea
\left[J_+,\Pi\right] &=& iQ\, , \qquad \left[J_-,Q\right]= -i\Pi\, , \nonumber\\
\left[J_3,\Pi\right]  &=& -\frac{1}{2}\Pi\, , \qquad \left[J_3, Q\right]
= \frac{1}{2}Q\,,\nonumber\\
\left[F,\Pi\right]  &=& -\frac{1}{2}\Pi \, , \qquad \left[F, Q\right] = -\frac{1}{2}Q\,, \lb{FPIQ}
\eea
\bea
\left[J_-,\Pi^\dagger \right] &=& iQ^\dagger\, , \qquad
\left[J_+,Q^\dagger \right]= -i\Pi^\dagger\, , \nonumber\\
\left[J_3,\Pi^\dagger\right]  &=& \frac{1}{2}\Pi^\dagger\, , \qquad
\left[J_3, Q^\dagger\right] = -\frac{1}{2}Q^\dagger\,,\nonumber\\
\left[F,\Pi^\dagger \right]  &=& \frac{1}{2}\Pi^\dagger \, ,
\qquad \left[F, Q^\dagger\right] = \frac{1}{2}Q^\dagger\, , \lb{FPQ2}
\eea
\bea
\left\{\Pi,\Pi^\dagger\right\} &=& -J_3 + F \, , \qquad
\left\{Q,Q^\dagger\right\} = J_3 +F\,,  \nonumber\\
\left\{\Pi, Q^\dagger\right\} &=& iJ_- \, , \qquad
\left\{\Pi^\dagger,Q\right\} = -iJ_+\, . \lb{PiQ}
\eea

The quadratic Casimir operator of the superalgebra $\textit{su}(2|1)$,
\be
C_2 =-\frac{1}{2}\{J_+, J_-\}-J_3^2 +F^2+\frac{1}{2}[Q, Q^\dag]+\frac{1}{2}[\Pi, \Pi^\dag],
\ee
is related to the Hamiltonian of the $n{=}2$ model as
\be
C_2 =H+4\kappa ^2 = H' - 2\kappa + 4\kappa^2. \label{n2c2}
\ee

One can also define a third-order Casimir operator:
\be
C_3 = \frac{i}{2}J_+[Q^\dagger,\Pi]-\frac{i}{2}[\Pi^\dagger, Q]J_-+\frac{1}{2}J_3([Q,Q^\dagger]-[\Pi,\Pi^\dagger])-\nn
\ee
\be
-\frac{1}{2}F([\Pi,\Pi^\dagger]+[Q,Q^\dagger])+2C_2F-\Pi^\dagger\Pi -QQ^\dagger\,.
\ee
In the $n{=}2$ model it can be represented as
\be
C_3=6\kappa H'+2\kappa(2\kappa -1)(4\kappa -1)\,.\label{c3n3}
\ee

\setcounter{equation}{0}

\section{The energy spectrum and wave functions}

In this section we turn to the study of the quantum $SU(n|1)/U(n)$ model.
We construct the complete set of the wave superfunctions and find the corresponding energy levels.
We also obtain the realization of the $SU(n|1)$ symmetry group on the wave superfunctions.

\subsection{The spectrum}
In this subsection we construct the complete set of wave superfunctions for the $SU(n)$ singlet sector of the full space of
quantum states. The corresponding superfunctions carry no external $SU(n)$ indices, but possess the fixed $B$
charge $-2\kappa \neq 0$, in accord with the explicit structure of the quantum generators \p{Qkvan}, \p{JBkvan}. Possible
wave functions with non-zero external $SU(n)$ spins form a subspace orthogonal to the $SU(n)$ singlet one
with respect to the inner product to be defined below. A similar situation occurs in the case of super Landau
models associated with the supersphere $SU(2|1)/U(1|1)$ \cite{BCIMT} where one can consider only those wave functions which
are singlets of the semi-simple part $SU(1|1) \subset U(1|1)\,$.

To proceed, we will need two important properties:
\be
(g^{-1})^i_j \left[\nabla _i^{(\kappa )} ,\; g^j_k\right]=\frac{(1-n)\bar {\xi }_k }{1- \xi \cdot\bar {\xi }} \lb{Prop1}
\ee
and
\be
\nabla _{[i}^{(\kappa )} \left(g^j_{k]} \Phi\right) =g^j_{[k} \nabla _{i]}^{(k+2)} \Phi\,, \lb{Prop2}
\ee
where $\Phi$ is an arbitrary superfunction, $\Phi = \Phi (\xi, \bar\xi)\,$, and
square brackets denote antisymmetrization of indices (with the factor $1/n!$). These relations can be proved
using the definitions \p{Metr} and \p{longd}.

We will deal with the shifted Hamiltonian $H'$ defined in \p{QH1}.
The lowest (vacuum) Landau level (LLL) wave function $\Psi_0$ corresponds to the zero eigenvalue of $H'$
and is defined by the same chirality condition as in ref. \cite{Odd}
\be
\nabla^{(\kappa)\bar {j}}\Psi _0=0 \quad \rightarrow \quad \Psi_0 = \frac{1}{(1 -\xi\cdot\bar\xi)^\kappa}\,\Omega_0(\xi)\,, \lb{LLL}
\ee
where the analytic wave function  is defined by the expansion
\be
\Omega_0(\xi)=c^0+\xi ^i\,c_i^0+ \ldots + \xi ^{i_1}\cdots\xi^ {i_n}\,c^0_{i_1\cdots i_n}\,.\lb{LLLan}
\ee

The wave superfunctions corresponding to the excited Landau levels  are constructed by acting
of the covariant derivatives $\nabla^{(\kappa')}$ on the chiral superfunctions. The latter should carry
the appropriate external $SU(n)$ indices in order to ensure the full
wave functions to be $SU(n)$ singlets.

The first LL wave superfunction is defined as
\be
\Psi_1 =\nabla_k^{(\kappa+1-n)} \Phi ^k,    \label{Psi_1}
\ee
where $\Phi ^k$ is the chiral superfunction in the fundamental representation of $U(n)$:
\be
\nabla^{(\kappa )\bar {j}}\Phi^k=0 \quad \rightarrow \quad \Phi^k  = \frac{1}{(1 -\xi\cdot\bar\xi)^\kappa}\,\Omega^k(\xi)\,. \lb{1LL}
\ee
Using \p{Prop1}, it is easy to check that $\Psi_1$ is the eigenfunction of $H'$,
\be
H'\,\Psi_1 = (2\kappa +1-n)\,\Psi_1\,.
\ee

The second LL wave superfunction is defined by
\be
\Psi _2 =\nabla _i^{(\kappa +1-n)} \nabla _k^{(\kappa +3-n)} \Phi ^{[ik]}, \label{Psi_2}
\ee
where $\Phi ^{[ik]}$ is a chiral superfield (it is expressed through the holomorphic reduced wave superfunction
$\Omega^{[ik]}(\xi)$ in the same way as in \p{LLL} and \p{1LL}).
The reason why the chiral superfunctions should belong to the irreducible
$U(n)$ representations constructed by antisymmetrizing the indices in the fundamental representation
will be explained in the next subsection.

Using \p{Prop1} and \p{Prop2}, one may verify that
\be
H'\,\Psi_2 = 2(2\kappa +2-n)\,\Psi_2\,.
\ee
In the $n{=}2$ case, when the indices take values $1$ and $2$, it is the last level in the spectrum,
because no non-zero higher-rank antisymmetric tensors can be defined.

For the same reason, in the general case of $n$-dimensional model the spectrum  terminates at the level $\ell = n$,
so we are left with the finite set of $n$ excited states. The wave superfunction for the $\ell\, ${\rm th} LL
is given by the expression
\be
\Psi _\ell =\nabla _{m_1 }^{(\kappa +1-n)} \nabla _{m_2 }^{(\kappa +3-n)}\cdots \nabla _{m_\ell }^{(\kappa +2\ell-1-n)}
\Phi ^{[m_1 m_2 \cdots m_\ell ]}\,,
\label{Psi-k}
\ee
where the reduced wave superfunction $\Phi ^{[m_1 m_2 \cdots m_\ell]}$ is chiral,
\be
\Phi ^{[m_1 m_2 \cdots m_\ell]} = \frac{1}{(1 - \xi\cdot \bar\xi)^\kappa}\,\Omega^{[m_1 m_2 \cdots m_\ell]}(\xi)\,. \lb{chirn}
\ee
The corresponding energy eigenvalue is
\be
E_\ell = \ell(\ell-n+2\kappa)\,. \lb{spectrum}
\ee
Sometimes it is convenient to use the equivalent representation for $\Psi_\ell$:
\bea
\Psi _\ell := \frac{1}{(1 - \xi\cdot \bar\xi)^\kappa}\,\hat{\Psi} _\ell\,, \quad \hat{\Psi} _\ell=
\nabla _{m_1 }^{(2\kappa +1-n)} \nabla _{m_2 }^{(2\kappa +3-n)}\cdots
\nabla _{m_\ell }^{(2\kappa +2\ell-1-n)}
\Omega^{[m_1 m_2 \cdots m_\ell ]}\,.
\label{Psi-k1}
\eea

It is natural to require that the energies of the excited Landau levels are not negative and exceed (or at least are not less than)
the energy of LLL. Therefore, in what follows we will consider only
the options when the WZ term strength is restricted to the values
\be
    \kappa \geq (n-1)/2\,. \lb{restrKappa}
\ee

\indent

The obtained set of $n+1$ chiral superfunctions captures the whole spectrum of the model in the $SU(n)$ singlet sector.
To prove this, we should check that any $SU(n)$ singlet superfunction can be expressed as a linear superposition of
the wave superfunctions of $n+1$ Landau levels.
The total number of independent functions of $n$ complex Grassmann variables is equal to $2^{2n}\,$.
The total number of independent coefficients in the $\xi^i$ expansion of holomorphic superfunction corresponding
to the level $\ell$ is $2^n{n\choose \ell}\,$. Different levels possess independent wave superfunctions.
So the total number of independent coefficients is
\be
2^n\sum _{\ell=0}^{n}{n\choose \ell} =2^{2n}.
\ee
Thus the constructed set of wave superfunctions indeed spans
the full $SU(n)$ singlet Hilbert space.

In the remainder of this subsection we will briefly discuss the case with $\kappa < 0$. Consider, instead of the $H'$
eigenvalues \p{spectrum}, those of the full hamiltonian $H = H' - \kappa n$:
\be
E_\ell \;\Rightarrow \; E_\ell = \ell(\ell-n+2\kappa) - \kappa n\,. \lb{spectrum1}
\ee
Now we assume that $\kappa = -|\kappa|$ and redefine the level number $\ell$ as
\be
\ell = -\ell' + n \,.\lb{redefell}
\ee
In terms of $\ell'$, the spectrum \p{spectrum1} for $\kappa < 0$ becomes
\be
E_{\ell'}^{(\kappa < 0)} = \ell'(\ell'-n+2|\kappa|) - |\kappa| n\,.\lb{spectrum2}
\ee
This formula coincides with that for $\kappa > 0\,$,
\be
E_{\ell}^{(\kappa > 0)} = \ell(\ell-n+2|\kappa|) - |\kappa| n\,,\lb{spectrum3}
\ee
modulo the substitution $\ell \rightarrow \ell'\,$. Thus, for $\kappa < 0\,$, the tower of the LL states is reversed:
the highest LL with $\ell =n$ becomes the LLL with $\ell' = 0$, while the LLL with $\ell = 0$ becomes the highest LL
with $\ell' = n\,$. In order to have, in both cases,
the excited LL energies to be not less than the LLL energy, one needs to impose the following general condition
\be
    |\kappa| \geq (n-1)/2\,. \lb{restrKappa}
\ee

The $\kappa <0$ LLL wave superfunction $\tilde{\Psi}_{\ell' =0} := {\Psi}_{\ell =n}$ can be checked to satisfy
the anti-chirality condition\footnote{The proof of this property is rather tricky.
One rewrites $\nabla^{(\kappa)}_i$ as $\nabla^{(\kappa)}_i = \partial_i + \kappa B_i\,,$ $B_i =
\bar{\xi}_i/(1 - \xi\cdot\bar\xi)\,$, and repeatedly
uses the identity $\partial_k B_i = B_k B_i\,$ to represent
the product of $n$ covariant derivatives in $\hat{\Psi}_{\ell = n}\,$, eq. \p{Psi-k1}, as a
differential operator in $\partial_i$ of the $n$-th order, with the coefficients
being monomials in $B_k\,$. To show that $\nabla^{(2\kappa)}_i\hat{\Psi}_{\ell = n} =0\,$,
one has to take into account the total antisymmetry in the indices $m_1 \ldots m_n$ and to make use of
the proper cyclic identities.}
\be
\nabla^{(\kappa)}_j \tilde{\Psi}_{\ell' =0} = 0 \quad \rightarrow \quad \tilde{\Psi}_{\ell' =0} =
(1 - \xi\cdot \bar\xi)^{-|\kappa|}\,\Omega_0(\bar\xi)\,,
\ee
whereas all other ones, up to $ \tilde{\Psi}_{\ell' =n}:= {\Psi}_{\ell =0}$, are obtained through the successive action
of the proper anti-holomorphic covariant derivatives on the anti-chiral superfunctions $(1 - \xi\cdot \bar\xi)^{-|\kappa|}$
$\Omega^{[i_1\ldots i_{\ell'}]}(\bar\xi)\,$.
Passing to the complex-conjugate
set of the wave superfunctions, $\tilde{\Psi}_{\ell'} \rightarrow \tilde{\Psi}_{\ell'}^\star\,$  takes us back to the holomorphic
representation, i.e. to the same picture as in the
$\kappa > 0$ case (with replacing  $\kappa \rightarrow |\kappa|\,$ everywhere). Thus, without loss of generality, we can
basically limit our study to the $\kappa > 0$ option.

\subsection{Transformation properties}
The $SU(n|1)$ transformation law of the wave function for any LL,
\be
\delta\Psi (\xi,\bar\xi )=\left(\epsilon \cdot Q+ \bar\epsilon \cdot Q^\dagger \right)\Psi (\xi,\bar\xi )\,, \lb{GenTran}
\ee
is fully specified by the form of the quantum supercharges:
\be
Q_i=-\partial _i+\bar \xi _i\bar\xi _j\partial ^{\bar j}+\kappa \bar \xi _i\,,  \label{Q}
\ee
\be
Q^{\dagger i}=-\partial ^{\bar i}-\xi ^i\xi ^j\partial _j+\kappa \xi ^i\,.
\ee
The bosonic transformations are contained in the closure of these odd ones.

Sometimes it is more convenient to deal with the equivalent {\it passive} form of the same $SU(n|1)$ transformation:
\be
\delta^* \Psi (\xi, \bar\xi) \simeq \Psi' (\xi', \bar\xi') - \Psi (\xi, \bar\xi)
=\kappa\left(\epsilon \cdot \bar{\xi}+ \bar{\epsilon}\cdot \xi \right)\Psi(\xi, \bar\xi)\,.  \label{supop}
\ee
Note that this transformation law indicates that the wave superfunctions cannot be real unless $\kappa = 0\,$.

Given the transformation law of the full wave superfunction for the level $\ell$, one can restore
the transformation rules of the reduced chiral superfunctions defined in the previous subsection.
To this end, one should take into account the ``passive'' transformation properties of the semi-covariant derivatives
\bea
&& \delta^* \nabla^{(\kappa')}_j = - (\bar\epsilon\cdot\xi)\,\nabla^{(\kappa')}_j + \bar\epsilon_j\xi^i\,\nabla^{(\kappa')}_i
+ \kappa'\, \bar\epsilon_j\,, \nn
&&\delta^* \nabla^{(\kappa')\bar j} = (\epsilon\cdot \bar\xi)\,\nabla^{(\kappa')\bar j}
- \epsilon^j\bar\xi_i\,\nabla^{(\kappa')\bar i} + \kappa'\,\epsilon^j\,. \lb{Trannabla}
\eea
Then we consider the transformation law of some  wave function $\Psi_\ell$ with $\ell\ge 2$ and require that
the corresponding transformation of $\Psi_\ell$ is given by the ``passive'' form of \p{GenTran}, i.e. by \p{supop}.
We find, first,  that $\Phi^{i_1\cdots i_\ell}$ should necessarily be fully antisymmetric in its $SU(n)$ indices,
$\Phi^{i_1\cdots i_\ell} = \Phi ^{[i_1\cdots i_\ell]}\,$, and, second, that  the transformation law of $\Phi^{[i_1\cdots i_\ell]}$
should be:
\be
\delta^*\Phi^{[i_1\cdots i_\ell]}=\kappa (\epsilon \cdot\bar{\xi})\,\Phi ^{[i_1\cdots i_\ell]}
+(\kappa +\ell)(\bar{\epsilon}\cdot\xi)\, \Phi^{[i_1 \cdots i_\ell]}+ \xi^{i_1}\bar{\epsilon}_j\,\Phi^{[j \cdots i_\ell]}+ \cdots +
\xi^{i_\ell}\bar{\epsilon}_j\,\Phi^{[i_1\cdots i_{\ell-1} j]}.\lb{Tranhol}
\ee
The chirality conditions are automatically covariant. For the first LL function $\Phi^i$ we have
the same transformation law, for the $SU(n)$ singlet LLL function $\Psi_0$
the $SU(n)$ rotation terms are obviously absent and the transformation law coincides with \p{supop}.
For the holomorphic wave superfunction
\be
\Omega^{[i_1\cdots i_\ell]}(\xi)  = (1 - \xi\cdot \bar\xi)^\kappa\,\Phi^{[i_1\cdots i_\ell]} \lb{Manchir}
\ee
the weight factors are properly combined in such a way  that the holomorphy property is preserved:
\be
\delta^*\Omega^{[i_1\cdots i_\ell]}(\xi) = (2\kappa +\ell)(\bar{\epsilon}\cdot\xi) \,\Omega^{[i_1 \cdots i_\ell]}(\xi) + \ldots\;,
\ee
where ``dots''  stand for the holomorphic $SU(n)$ rotations, which are the same as in \p{Tranhol}. This transformation
law is valid for any $\ell \geq 0\,$.

\subsection{$n{=}2$}\label{n=2}
Here we consider in more detail the $n{=}2$ case, which corresponds to the $SU(2|1)/U(2)$ model.
In this case we can lower and raise the $SU(2)$ indices with the help of skew-symmetric symbols,
\beq
\varepsilon_{ik}\,,\;\; \varepsilon^{ik}\,,\;\;
\varepsilon^{ik}\varepsilon_{kj} = \delta^i_j\,,\;\; \varepsilon_{12} = \varepsilon^{21} = 1\,.
\eeq
The holomorphic LLL superfunction $\Omega_0(\xi)$ has the following $\xi$-expansion:
\be
\Omega _0(\xi)=c^0+ \xi ^i c^0_i+ \frac{1}{2}\xi ^{i_1}\xi ^{i_2}\varepsilon_{i_1i_2}c_1\,, \lb{defc0c1}
\ee
where $c_0$ and $c_1$ are $SU(2)$ singlets.
{}From the transformation law
\be
\delta \Omega_0 = 2\kappa (\bar\epsilon \cdot \xi) \Omega_0 - [\epsilon^i + (\bar\epsilon \cdot \xi)\xi^i]\partial_i\Omega_0
\ee
we derive the following transformations of the component wave functions
\bea
\delta c^0= -\epsilon ^ic^0_i \,,\quad
\delta c^0_i= - 2\kappa \bar\epsilon_i c^0- \epsilon_i c_1\,, \quad
\delta c_1=(2\kappa -1)\bar\epsilon^k c^{0}_k\,.\lb{tr0}
\eea
Here, the complex conjugation rule for $\epsilon_i$ is
\be
    \left(\epsilon^i\right)^* = \bar\epsilon_i\,, \quad \left(\epsilon_i\right)^* = -\bar\epsilon^i\,.
\ee

Next, we consider the first level wave superfunction:
\be
\Omega ^{[k]}(\xi)=c^{[k]}+\xi ^j c^{[k]}_j+ \frac{1}{2}\xi ^{j_1}\xi ^{j_2}\varepsilon_{j_1j_2}c^{[k]}_2\,. \label{phi11}
\ee
It transforms according to the rule
\be
\delta \Omega^{[k]} = (2\kappa +1)(\bar\epsilon \cdot \xi) \Omega^{[k]} + \xi^k\bar\epsilon_i\,\Omega^{[i]}
- [\epsilon^i + (\bar\epsilon \cdot \xi)\xi^i]\partial_i\Omega^{[k]}\,,
\ee
which implies the following component transformation laws:
\bea
\delta c^{[k]} = -\epsilon ^i\, c_i^{[k]},\;
\delta c^{[k]}_j = -\epsilon _jc_2^{[k]} -(2\kappa
+1)\bar\epsilon _jc^{[k]}+\delta ^k_j\bar\epsilon _ic^{[i]}, \;
\delta c_2^{[k]}=\varepsilon ^{kj}\bar\epsilon_i c^{[i]}_{j}
+2\kappa \bar\epsilon^{j} c^{[k]}_{j}.\lb{tr1}
\eea

Finally, consider the second-level wave superfunction:
\be
\Omega ^{[k_1k_2]}(\xi)=c^{[k_1k_2]}+\xi ^jc^{[k_1k_2]}_j+\xi ^{j_1}\xi ^{j_2}\,c^{[k_1k_2]}_{j_1j_2}\,. \label{phi112}
\ee
Introducing
\be
A=\varepsilon _{i_1i_2}c^{[i_1i_2]},\quad B=\varepsilon ^{k_1k_2}\varepsilon _{i_1i_2}c^{[i_1i_2]}_{k_1k_2},\quad
F_k=\varepsilon _{i_1i_2}c_k^{[i_1i_2]}\,,
\ee
we find
\bea
\delta A=-\epsilon ^kF_k\,, \quad
\delta B= -2\kappa \bar\epsilon ^{k}F_{k}\,, \quad
\delta F_k=\epsilon_k B-(1+2\kappa)\bar\epsilon _kA\,.\lb{tr2}
\eea
Note that these transformations can be brought precisely into the form \p{tr0} after redefinition $B \rightarrow -B, \;
2\kappa \rightarrow 2\kappa -1\,$. This means that the $\ell = 0$ and $\ell =2$ wave superfunctions constitute isomorphic $SU(2|1)$
multiplets.

It is appropriate here to give the relevant values of the $su(2|1)$ Casimirs $C_2$ and $C_3$ defined in \p{n2c2}
and \p{c3n3}.

The eigenvalues of the Casimir operators on the LLL state are
\be
    C_2 =2\kappa\left(2\kappa -1\right)\,,\quad C_3=2\kappa(2\kappa -1)(4\kappa -1)\,.\lb{C230}
\ee
On the first LL the Casimir operators become
\be
    C_2 =\left(2\kappa +1\right)\left(2\kappa -1\right)\,, \quad C_3=4\kappa(2\kappa -1)(2\kappa +1)\,.\lb{C231}
\ee
Finally, their eigenvalues on the second level are
\be
    C_2 =2\kappa\left(2\kappa +1\right)\,,\quad C_3=2\kappa(2\kappa +1)(4\kappa +1)\,.\lb{C232}
\ee
Note that the replacement $2\kappa \rightarrow 2\kappa -1$ in \p{C232} yields just \p{C230},
in accord with the remark after eqs. \p{tr2}.

The spectrum of Casimir operators for the finite-dimensional representations of $SU(2|1)$ was studied in \cite{Casim,Casim1}.
These representations are characterized by some positive number $\lambda$ (``highest weight'') which can be half-integer or integer
and an arbitrary additional real number $\beta$  which is related to the eigenvalues of the generator $F$ (``baryon charge'').
The values \p{C230} - \p{C232} can be uniformly written in the generic form given in \cite{Casim} as
\be
C_2 = (\beta^2 - \lambda^2)\,, \quad C_3 = 2\beta(\beta^2 - \lambda^2) = 2\beta C_2\,,\lb{Casimunif}
\ee
with
\bea
\underline{\rm LLL}: \qquad &\lambda& = \frac{1}{2}\,, \quad \beta = \frac{4\kappa - 1}{2}\,, \lb{LLL0} \\
\underline{\rm 1st \;LL}: \qquad &\lambda& = 1\,, \quad \beta =  2\kappa\,, \lb{LL1} \\
\underline{\rm 2nd \;LL}: \qquad &\lambda & = \frac{1}{2}\,, \quad \beta = \frac{4\kappa + 1}{2}\,. \lb{LL2}
\eea
Note that our $C_2$ and $C_3$ were defined to have the opposite sign to those in \cite{Casim,Casim1} and
$C_3$ also differs by the factor 2. We also took into account that $\kappa \geq 1/2$ in our case.
The isospins and $F$-charges of the component wave functions are expressed through the appropriate quantum numbers
$\lambda$ and $\beta$ in full agreement with the general formulas of \cite{Casim}.

While at $\kappa > 1/2$ we deal with what is called ``typical'' $SU(2|2)$ representations (both Casimirs are non-zero),
at the special value $\kappa =1/2$ both Casimirs are zero for the LLL and 1st LL multiplets. So in this case the latter
belong to the so called ``atypical'' $SU(2|2)$ representations. In accord with the consideration in \cite{Casim1}, they are
not completely irreducible: they contain invariant subspaces the quotients over which, in turn, yield some further irreducible
representations. As is seen from the transformation properties \p{tr0}, \p{tr1}, at $2\kappa =1$ the component $c_1$ of
$\Omega_0$ is $SU(2|1)$ singlet and the subset $(c^{[k]}_k, c^{[k]}_2)$ in $\Omega_1$ also forms
a closed $SU(2|1)$ multiplet.

In the alternative $\kappa < 0$ case the eigenvalues of the quadratic Casimir  are
\bea
    C_2 &=&2|\kappa |\left(2|\kappa | +1\right)\,,\qquad \quad \;\, \ell=0 \quad (\ell' =2)\,,\nn
    C_2 &=&\left(2|\kappa | -1\right)\left(2|\kappa | +1\right)\,,\quad \ell=1 \quad (\ell' =1)\,,\nn
    C_2 &=&2|\kappa |\left(2|\kappa | -1\right)\,,\qquad \quad \; \, \ell=2 \quad (\ell' =0)\,.
\eea
These values are not negative for $|\kappa |\geq 1/2\,$, in accord with the general condition \p{restrKappa}.
\setcounter{equation}{0}
\section{$SU(n|1)$ invariant norms}
\subsection{General case}
The $SU(n|1)$ invariant Berezin integral is defined as \cite{Odd}
\be
\int d\mu=\int d\mu _0 \left[1-(\xi\cdot\bar{\xi})\right]^{n-1}, \label{invm}
\ee
where
\be
\int d \mu_0=\prod  \partial _i\partial ^{\bar{i}}\,.
\ee
Using this integration measure, we can define the inner product
on the Hilbert space of wave superfunctions:
\be
\langle\Psi\big|\Omega\rangle=\int d\mu\Psi^{\star}\Omega\,.   \label{invnorm}
\ee
It is manifestly $SU(n|1)$ invariant, taking into account the invariance of the measure $d\mu$
and the fact that the weight factor in the general transformation law \p{supop} is imaginary.

To express the norms in terms of the reduced chiral superfunctions $\Phi^{[k_1\cdots k_n]}$, defined in \p{Psi-k}, we will need
the following rule of integration by parts for two superfunctions, $\Theta$ and $\Phi$:
\be
\int d\mu \Theta (\nabla _k^{(\kappa)}\Phi)^{\star}= (-1)^{P(\Theta) + P(\Phi)} \int d\mu (\nabla ^{(\kappa +n-1) \bar{k}}
\Theta) \Phi^{\star}\,, \label{oddphi^k}
\ee
where $P(\Theta)$ and $P(\Phi)$ are Grassmann parities of the superfunctions. We will also employ the identity
\be
\nabla^{(\kappa +2)[\bar{i}}g^{k]}_j=g^{[k}_j\nabla^{(\kappa)\bar{j}]}.
\ee

Using these rules, together with the anticommutation relations \p{ant}, and the anti-chirality condition
$$
\nabla^{-(\kappa)}_j(\Phi^{[k_1\cdots k_n]})^* = 0\,,
$$
one can express the norm of the full wave superfunction $\Psi_\ell$ for the $\ell$-th LL in terms of the chiral wave functions as
\be
\big|\big|\Psi_\ell\big|\big| ^2=\frac{\ell!(2\kappa -n+2\ell -1)!}{(2\kappa -n+\ell -1)!}
\int d\mu g^{m_1}_{i_1}\cdots g^{m_\ell}_{i_\ell}(\Phi^{[m_1\cdots m_\ell]})^{\star}\Phi^{[i_1\cdots i_\ell]}\,. \lb{NormPhi}
\ee
It is also straightforward to show that the wave superfunctions associated with different Landau levels are mutually orthogonal.

Expressing $\Phi^{[k_1\cdots k_n]}$ through holomorphic wave functions by eq. \p{Manchir},
$$
\Phi^{[k_1\cdots k_n]} = (1 - \xi\cdot \bar\xi)^{-\kappa}\Omega^{[k_1\cdots k_n]}(\xi)\,,
$$
one can perform the $\xi$ integration in \p{NormPhi} and obtain the norms written in terms of the
coefficients in the $\xi$ expansion of the $\Omega^{[k_1\cdots k_n]}(\xi)\,$. As an illustration, we present
this final form of the norm of the LLL wave superfunction, with the $\xi$-expansion defined
in \p{LLLan}:
\be
\big|\big|\Psi_0\big|\big| ^2=(-1)^n\sum_{k=0}^n{n-2\kappa -1 \choose k}k!(n-k)!
\bar{c}^{0i_1\cdots i_{n-k}}c_{i_1\cdots i_{n-k}}^0\,.\lb{LLLnorm}
\ee

In the next subsection, as an instructive example,  we will give the explicit expressions for norms of all
three LLL states of the $n{=}2$ model.

As should be clear from the expression \p{LLLnorm}, there are values of $\kappa$ for which the squared norms
are negative, the same is true for the norms of higher LL. This  situation is typical for
quantum-mechanical systems with Grassmann-odd target space coordinates  \cite{Odd}.
In the next Section,  we analyze this issue in some detail on the $n{=}2$ example. The way to make all norms
positive-definite  is to modify the inner product by introducing some
metric operator on the Hilbert space (like in all other known examples of super Landau models).
This operator proves to be especially simple in the planar limit (section 7).

\subsection{Norms for the $SU(2|1)/U(2)$ model}
Here we specialize to the $n{=}2$ case and present the explicit expressions for the ``naive'' norms, using
the general formulas of the previous subsection.

The norm of the vacuum wave superfunction is given by
\be
||\Psi _0||^2=\bar c_1 c_1+(1-2\kappa)\bar c^{0i}c^0_i+2\kappa (2\kappa -1)\bar c^0 c^0\,,\label{Norm1}
\ee
where the component wave functions were defined in \p{defc0c1}. The first excited level is described
by the wave superfunction with the norm
\be
||\Psi _1||^2=(2\kappa -1)\left[(2\kappa +1)(2\kappa -1)\bar c_{[k]}c^{[k]}+(1-2\kappa )\bar c^i_{[k]}c^{[k]}_i-\bar c_{[k]}^kc^{[i]}_i
+\bar c_{2[k]}c_2^{[k]}\right]. \label{Norm3}
\ee
The second level wave superfunction has the norm
\be
||\Psi _2||^2=2\kappa (2\kappa +1)\left[2\kappa (2\kappa +1)\bar A A+\bar B B-2\kappa\bar F^kF_k\right].\label{Norm2}
\ee

It is straightforward to check that these norms are invariant under the transformations \p{tr0} - \p{tr2}.
Also we observe that these norms are not positive-definite. In the next section we will see in detail how this unwanted
property can be cured. At $2\kappa =1$ there are zero norms for the LLL and the 1st LL wave functions. In this case it
is natural to define the physical Hilbert space as a quotient over the subspace of zero-norm states, so it is spanned by the
LLL $SU(2|1)$ singlet ``wave function'' $c_1$ and 4 wave functions $(A, B, F^k)$ of the second LL.

\section{Unitary norms for $n{=}2$ and hidden $SU(2|2)$ \break symmetry}
\subsection{Redefining the inner product}
All norms for the case $n{=}2$ can be made positive by modifying the inner product in the Hilbert space, like in the cases
worked out in \cite{BCIMT}:
\be
\langle\langle\Psi |\Phi\rangle\rangle =\int d\mu\Psi ^\star G\Phi\, , \lb{Impr}
\ee
where $G$ is a metric operator on Hilbert space. As was already mentioned in subsection 4.1, we will consider only
the case $|\kappa| \geq 1/2$, because only in this case the energies of the excited Landau levels are non-negative.
For the case $\kappa \geq 1/2\,$, we choose the metric operator to be
\be
G=1-4\left(2F+4\kappa +\ell\right)+2\left(2F+4\kappa +\ell\right)^2\,.\lb{Ggen}
\ee
It satisfies the conditions
\be
G^2=1\,,\qquad \left[H, G\right] =0\,,
\ee
which mean that the metric operator just alters the sign in front of all negative terms in the
expressions for the norms of the wave superfunctions. Accordingly, the positive norms of the wave superfunctions are given by
\bea
&&||\Psi _0||^2=\bar c_1 c_1+(2\kappa -1)\bar c^{0i}c^0_i+2\kappa (2\kappa -1)\bar c^0 c^0\,,\nn
&&||\Psi _1||^2=(2\kappa -1)\left[(2\kappa +1)(2\kappa -1)\bar c_{[k]}c^{[k]}+(2\kappa -1)\bar c^i_{[k]}c^{[k]}_i
+\bar c_{[k]}^kc^{[i]}_i
+\bar c_{2[k]}c_2^{[k]}\right],\nn
&&||\Psi _2||^2=2\kappa (2\kappa +1)\left[2\kappa (2\kappa +1)\bar A A+\bar B B+2\kappa\bar F^kF_k\right].\label{pnorms}
\eea
With $\kappa=1/2\,$, the norms become
\bea
&&||\Psi _0||^2=\bar c_1 c_1\,,\qquad ||\Psi _1||^2=0\,,\nn
&&||\Psi _2||^2=2\kappa (2\kappa +1)\left[2\kappa (2\kappa +1)\bar A A+\bar B B+2\kappa\bar F^kF_k\right].\lb{0norms}
\eea
As was already mentioned, the Hilbert space in this special case, obtained as  quotient over the zero-norm states,
involves only two physical states, one corresponding to LLL
with unbroken $SU(2|1)$ symmetry and another one corresponding to the second LL.\\
\indent Now, let ${\cal O}$ be some operator that commutes with the Hamiltonian, and hence generates some symmetry
of the model, and let ${\cal O}^\dagger$ be its hermitian conjugated operator with respect
to the `naive' inner product \p{invnorm}. Then its hermitian conjugate with respect
to the `improved' product \p{Impr} is given by
\be
{\cal O}^\ddagger \equiv G {\cal O}^\dagger G ={\cal O}^\dagger + G\left[{\cal O}^\dagger ,G\right].\label{conj}
\ee
To find a new conjugation for the $SU(2|1)$ generators, we need to know their realization on the analytic
wave functions $\Omega(\xi)$:
\bea
&&\Pi =-\partial _1\,,\qquad \Pi^{\dagger}=-\xi^1\xi^2\partial_2+\xi^1\left(2\kappa +\frac{\ell}{2}-\hat{B}_3\right)-\xi^2  \hat B_+\,,\nn
&&Q=-\partial _2\,, \qquad Q^{\dagger}=-\xi^2\xi^1\partial_1+\xi^2\left(2\kappa +\frac{\ell}{2}+\hat{B}_3\right)-\xi^1\hat B_-\,,\nn
&&J_+ =-i\xi ^1\partial _2+i\hat B_+\,,\qquad J_- =i\xi ^2\partial _1-i\hat B_-\,, \nn
&&J_3 =\frac{1}{2}(\xi ^1\partial _1 -\xi ^2\partial _2 )-\hat{B}_3\,, \nn
&&F=\frac{1}{2}(\xi ^1\partial _1 +\xi ^2\partial _2 -4\kappa -\ell)\,.\label{diff}
\eea
The matrix parts $\hat B$ of the $SU(2)$ generators satisfy, on their own,  the $su(2)$ commutation relations,
\be
    \left[\hat B_+,\hat B_-\right] = -2\hat B_3\, , \qquad \left [\hat B_3,\hat B_\pm\right] =\mp \hat B_\pm\, .
\ee
They can take non-vanishing values only when applied to the first-level analytic wave functions $\Omega^{[k]}$ which form
a doublet with respect to the external index:
\bea
    &&\hat B_+ \Omega^{[2]}=\Omega^{[1]}\,,\qquad \hat B_+ \Omega^{[1]}=0 \,,\nn
    &&\hat B_- \Omega^{[1]}=\Omega^{[2]}\,,\qquad \hat B_- \Omega^{[2]}=0 \,,\nn
    &&\hat B_3 \Omega^{[1]}=\frac{1}{2}\Omega^{[1]}\,,\quad\;\; \hat B_3 \Omega^{[2]}=-\frac{1}{2}\Omega^{[2]} \,.
\eea

In the holomorphic realization, the differential ``metric'' operator $G$ defined in \p{Ggen} takes the form
\be
G_{an}=1 - 2\xi^i\partial_{i} + 4\xi^1\xi^2\partial_{2}\partial_{1}=(1 - 2\xi^1\partial_{1})(1 - 2\xi^2\partial_{2})\, . \lb{Gn2}
\ee
One can check that $G_{an}$ anticommutes with the supercharges and commutes with
the bosonic  $SU(2)\times U(1)$ generators. Owing to these properties, we find
\be
Q^\ddagger =-Q^\dagger ,\qquad \Pi^\ddagger =-\Pi^\dagger .\label{GQ}
\ee

In \cite{BCIMT}, there was given a general definition of supercharge that commutes with $G$:
\be
    \tilde O = O + \frac{1}{2}\left[G, O\right]G. \lb{CommG}
\ee
However, in the model under consideration it identically vanishes. So one cannot define a modified $su(2|1)$ superalgebra
which would commute with the operator $G$. Therefore, the $su(2|1)$ transformation properties of the component wave functions
are slightly changed after passing to the new Hermitian conjugation.

\indent The odd $SU(2|1)$ transformations are now generated by
\bea
    &&\delta  \Psi\equiv \left( \epsilon Q+ \bar{\epsilon}Q^\ddagger\right)\Psi\,,
\eea
giving rise to the following modified transformation properties of the component wave functions:
\bea
\delta c^0= -\epsilon ^ic^0_i \,,\quad
\delta c^0_i= 2\kappa \bar\epsilon_i c^0- \epsilon_i c_1\,, \quad
\delta c_1=-(2\kappa -1)\bar\epsilon^k c^{0}_k\,,\label{s21l0}
\eea
\bea
\delta c^{[k]} = -\epsilon ^i\, c_i^{[k]},\;
\delta c^{[k]}_j = -\epsilon _jc_2^{[k]} +(2\kappa
+1)\bar\epsilon _jc^{[k]}-\delta ^k_j\bar\epsilon _ic^{[i]}, \;
\delta c_2^{[k]}=-\varepsilon ^{kj}\bar\epsilon_i c^{[i]}_{j}
-2\kappa \bar\epsilon^{j} c^{[k]}_{j},
\eea
\bea
\delta A=-\epsilon ^kF_k\,, \quad
\delta B= -2\kappa \bar\epsilon ^{k}F_{k}\,, \quad
\delta F_k=\epsilon_k B -(1+2\kappa)\bar\epsilon _k A\,.
\eea
These transformations are similar to \p{tr0}, \p{tr1}, \p{tr2}, the difference being the opposite sign before
the terms with $\bar \epsilon_i$.
The modified norms \p{pnorms} are invariant just under these transformations.
The set of the group generators is now $Q, \Pi, Q^\ddagger, \Pi^\ddagger, F, J_\pm, J_3$. Quadratic Casimir operator
for them is
\be
C_2 =-\frac{1}{2}\{J_+, J_-\}-J_3^2 +F^2-\frac{1}{2}[Q, Q^\ddagger]-\frac{1}{2}[\Pi, \Pi^\ddagger]\,.
\ee
It is related to the Hamiltonian by the same eq. \p{n2c2} and so takes the same values \p{C230} - \p{C232}  on the LL wave
superfunctions (the same is true for the 3d order Casimir \p{c3n3}).
In particular, for $\kappa=1/2\,$, Casimir operators vanish for the LLL and the 1st LL superfunctions, implying
these levels to carry atypical representations of $SU(2|1)\,$. It is worthwhile to note that, although the $su(2|1)$ algebra
\p{JJ} - \p{PiQ} changes its form after the replacement
$Q^\dagger, \Pi^\dagger \rightarrow
-Q^\ddagger, -\Pi^\ddagger$, the original form can be restored by passing (in the $SU(2)$ covariant notation \p{QQ}, \p{QbarQ})
to the new generators $\tilde{Q}_i = -\epsilon_{ik}{Q}^{\ddagger k}, \; \tilde{Q}{}^{\ddagger i} =
\epsilon^{ik}{Q}_k\,,\; \tilde{F} = - {F}\,$.

\subsection{$SU(2|2)$ symmetry}\label{6.2}
It was shown in \cite{BCIMT} that the quantum Hilbert space of the superflag Landau model carries hidden $SU(2|2)$ symmetry
which is, in a sense, an analog of the hidden worldline ${\cal N}=2$ supersymmetry of the superplane Landau model \cite{CIMT,SupLan}.
It turns out that this phenomenon of enhancing the original $SU(2|1)$ symmetry to $SU(2|2)$ at the quantum level persists as well
in the considered odd-coset super Landau model. Below we assume that $\kappa >1/2$, i.e. that the Casimir operators are non-vanishing
for all three LLs.

Because of the anticommutation property $\{G, \Pi\} = \{G, \Pi^\dagger\} = \{G, Q\} = \{G, Q^\dagger\} =0\,$, the method
of defining hidden supersymmetries applied in \cite{BCIMT} is not directly applicable to the present case. In particular,
just due to this property, the supercharges commuting with $G$ and defined by the  general formula \p{CommG} are identically
vanishing. Yet, we can achieve our goal, though in a distinct way.

We define
\bea
&&\Pi_G  \equiv \frac{1}{2}\left[\Pi, G\right]=\Pi G\,, \qquad\Pi_G^{\ddagger} \equiv -\Pi^\ddagger G\,,\nn
&&Q_G \equiv \frac{1}{2}\left[Q, G\right]=Q G \,,\qquad Q_G^{\ddagger} \equiv -Q^\ddagger G\,.
\eea
These operators can be used to generate the second $SU(2)$ algebra\footnote{Due to the anticommutativity  of the metric operator $G$
with $\Pi$ and $Q$ (and its commutativity with bosonic generators) the operators $\Pi_G, \Pi_G^\ddagger$ and $Q_G, Q_G^\ddagger$
form the same $su(2|1)$
algebra as $\Pi, \Pi^\ddagger$ and $Q, Q^\ddagger$ themselves, but they do not produce any obvious closed structure
together with the latter. This feature is in contrast with the construction in ref. \cite{BCIMT}, where just the generators
$\Pi_G, \Pi_G^\ddagger$ and $Q_G, Q_G^\ddagger$ extend $su(2|1)$ to $su(2|2)$.} as
\bea
    &&{\cal J_-}\equiv \frac{1}{2\sqrt{C_2}}\lbrace Q, \Pi_G \rbrace =-\frac{1}{2\sqrt{C_2}}\lbrace \Pi , Q_G \rbrace
    =\frac{1}{\sqrt{C_2}}\,Q\Pi G\,,\nn
    &&{\cal J_+}\equiv \frac{1}{2\sqrt{C_2}}\lbrace\Pi_G^\ddagger,Q^\ddagger \rbrace
    = -\frac{1}{2\sqrt{C_2}}\lbrace Q_G^\ddagger, \Pi^\ddagger \rbrace =\frac{1}{\sqrt{C_2}}\,\Pi^\ddagger Q^\ddagger G \,,\nn
    &&{\cal J}_{3} = \frac{1}{2}\left[{\cal J_+}, {\cal J_-}\right].
\eea
The Casimir operator $C_2$ was defined in \p{n2c2}. Thus we have two distinct $SU(2)$ algebras
\bea
    &&\left[J_+, J_-\right]= 2 J_3\,,\qquad \left[J_3, J_\pm\right]= \pm J_\pm\,,\nn
    &&\left[{\cal J_+}, {\cal J_-}\right]= 2{\cal J}_3\,,\qquad \left[{\cal J}_3, {\cal J}_\pm\right]= \pm{\cal J}_\pm\,,\label{alt3}
\eea
which can be checked to commute with each other. Explicitly, in the holomorphic realization,
these generators read
\bea
    &&J_+ =J_2^1 =\xi ^1\partial _2-\hat B_+\,,\qquad J_- =J_1^2 =\xi ^2\partial _1-\hat B_-\,, \nn
    &&J_3 =J_1^1 =-J_2^2 =\frac{1}{2}(\xi ^1\partial _1 -\xi ^2\partial _2 )-\hat{B}_3\,, \lb{BosOne} \\
    &&{\cal J_+}={\cal J}_1^2=\sqrt{C_2}\xi^1\xi^2\,,\qquad {\cal J_-}={\cal J}_2^1=\frac{1}{\sqrt{C_2}}\partial_{2}\partial_{1}\,,\nn
    &&{\cal J}_3={\cal J}_1^1 =-{\cal J}_2^2 = \frac{1}{2}\left(\xi^1\partial_{1}+\xi^2\partial_{2}-1\right).\lb{BosTwo}
\eea
Remind that the $SU(2)$ matrix operators $\hat{B}$ are non-vanishing only at the first LL ($\ell=1$)
and are zero for other levels. Then we can define the supercharges $S^a_i$ $(i= 1,2; a = 1,2)$ which
are doublets with respect to either of two $SU(2)$ groups:
\bea
    &&S_1^1=\Pi ,\qquad S_1^2\equiv \left[\Pi, {\cal J_+}\right]\,,\nn
    &&S_2^1=Q , \qquad S_2^2\equiv \left[Q, {\cal J_+}\right],\label{sch1} \\
&& \bar{S}_a^i := \left(S_i^a\right)^\ddagger\,.
\eea
Explicitly, these supercharges are
    \bea
    &&S_1^1=-\partial _1,\qquad S_1^2=-\sqrt{C_2}\xi^2 \,,\nn
    &&S_2^1=-\partial _2,\qquad S_2^2=+\sqrt{C_2}\xi^1\,,\lb{Odin}
\eea
\bea
    &&\bar{S}_1^1=\xi^1\xi^2\partial_2 -\xi^1\left(2\kappa +\frac{\ell}{2}-\hat{B}_3\right)+\xi^2  \hat B_+\,, \nn
    &&\bar{S}_1^2=\xi^2\xi^1\partial_1 -\xi^2\left(2\kappa +\frac{\ell}{2}+\hat{B}_3\right)+\xi^1\hat B_-\,, \nn
    &&\sqrt{C_2}\bar{S}_2^1 =+\left(1-\xi^1\partial_{1}\right)\partial_2 -\left(2\kappa +\frac{\ell}{2}-\hat{B}_3\right)\partial_2
    -\hat B_+\,\partial_1\,,\nn
    &&\sqrt{C_2}\bar{S}_2^2 =-\left(1-\xi^2\partial_{2}\right)\partial_1 +\left(2\kappa
    +\frac{\ell}{2}+\hat{B}_3\right)\partial_1+\hat B_-\,\partial_2\,.\lb{Two}
\eea
It is straightforward to check that they satisfy the (anti)commutation relations of
the superalgebra ${\it su}(2|2)$  with three central charges:
\bea
    &&\lbrace S_{i}^{a}, \bar{S}_b^j\rbrace =\delta_b^a J_i^j - \delta_i^j{\cal J}_b^a  +\delta_b^a \delta_i^j\left(2\kappa
    +\frac{\ell}{2}-\frac{1}{2}\right)\,,\nn
    &&\lbrace S_{i}^{a},S_j^b\rbrace =\varepsilon_{ij}\varepsilon^{ab}\sqrt{C_2}\,,\qquad \lbrace \bar{S}_a^{i},
    \bar{S}_b^j\rbrace =\varepsilon^{ij}\varepsilon_{ab}\sqrt{C_2}\,,\nn
    &&\left[S_i^a, {\cal J}_b^c\right]=\delta_b^a S_i^c -\frac{1}{2}\delta_b^c S_i^a\,,\qquad
    \left[S_i^a,  J_k^j\right]=\delta_i^j S_k^a -\frac{1}{2}\delta_k^j S_i^a\,. \label{alg1}
\eea

As an example, let us give how this $SU(2|2)$ symmetry is realized on the LLL (i.e. $\ell =0$) wave functions.
Denote $\theta^i, \bar\theta_i$ the parameters associated with the extra pair of fermionic generators, i.e. with $S^2_i, \bar S^i_2$.
Then the additional transformations are
\bea
&&\delta c^0=\frac{\sqrt{2\kappa-1}}{\sqrt{2\kappa}}\,\bar\theta^i c^0_i\,,\quad
\delta c_1=-\sqrt{(2\kappa-1)(2\kappa)}\,\theta^k c^{0}_k\,,\nn
&&\delta c^0_i=-\frac{\sqrt{2\kappa}}{\sqrt{2\kappa-1}}\, \bar\theta_i c_1 - \sqrt{(2\kappa-1)(2\kappa)}\,\theta_i c^0\,.
\eea
It is easy to check that they leave invariant the norm $||\Psi _0||$ in \p{pnorms}. Their bracket with
the $SU(2|1)$ transformations \p{s21l0} produces the second $SU(2)\,$, with respect to which $c^0$ and $c_1$ form a doublet.
The remaining components $c^0_i$ are singlets of the second $SU(2)\,$.
It is also easy to find the realization of the $\theta$-transformations on the $\ell =1$ and $\ell=2$
wave function multiplets.

According to \cite{Bn} (see also \cite{AFr}), the central charges can be combined into the 3-vector $\vec{C}$ as
    \be
    \vec{C}(\ell)=\left(2\kappa  +\ell /2-1/2, \sqrt{C_2(\ell)}, \sqrt{C_2(\ell)}\right) \equiv \left(C, P, K\right).\lb{Cvector}
    \ee
The norm of $\vec{C}$ defined as $ \vec{C}^2 = C^2 - PK$ is invariant under the $so(1,2)$ outer automorphisms of the $su(2|2)$
superalgebra \p{alg1} \cite{Bn}. Exploiting this $so(1,2)$ freedom, one can cast $\vec{C}\,$ in the form
    \be
    \vec{C} =\left(Z, 0, 0\right), \lb{newFr1}
    \ee
where
\be
Z= \frac{1}{2} \;\; {\rm for} \;\ell = 0,2\,, \quad {\rm and} \;\;  Z= 1 \;\; {\rm for} \;\ell = 1\,.
\ee
Explicitly, the $so(1,2)$ rotated supercharges for all three levels are
\bea
    &&\tilde{S}^a_i= \sqrt{2\kappa}\,S^a_i -\sqrt{2\kappa -1}\,\varepsilon^{ab}\varepsilon_{ij} \bar{S}^j_b\,,\qquad \qquad
    \qquad \;\, \ell=0\,, \nn
    &&\tilde{S}^a_i= \sqrt{2\kappa +1}\,S^a_i -\sqrt{2\kappa}\,\varepsilon^{ab}\varepsilon_{ij} \bar{S}^j_b\,,\qquad \qquad \qquad \;
    \,\ell=2\,, \nn
    &&\tilde{S}^a_i=\frac{1}{\sqrt{2}}\left[ \sqrt{2\kappa +1}\,S^a_i -\sqrt{2\kappa -1}\,\varepsilon^{ab}\varepsilon_{ij}
    \bar{S}^j_b\right],\qquad \ell=1\,. \lb{tildeS}
\eea
In the new frame \p{newFr1}, \p{tildeS} the (anti)commutation relations of ${\it su}(2|2)$ become
\bea
    &&\lbrace \tilde S_{i}^{a}, \tilde{\bar{S}}_b^j\rbrace =\delta_b^a J_i^j - \delta_i^j{\cal J}_b^a  +Z\delta_b^a \delta_i^j\,,\nn
    &&\lbrace \tilde S_{i}^{a}, \tilde S_j^b\rbrace =0\,,\qquad \lbrace \tilde{\bar{S}}_a^{i}, \tilde{\bar{S}}_b^j\rbrace =0\,,\nn
    &&\left[\tilde S_i^a, {\cal J}_b^c\right]=\delta_b^a \tilde S_i^c -\frac{1}{2}\delta_b^c \tilde S_i^a\,,
    \qquad\left[\tilde S_i^a,  J_k^j\right]=\delta_i^j \tilde S_k^a -\frac{1}{2}\delta_k^j \tilde S_i^a\,.\label{alg2}
\eea

With respect to the redefined $su(2|2)$ generators, the transformations of the LLL supermultiplet take the form
\bea
\delta c^0= -\frac{\rho^ic^0_i }{\sqrt{2\kappa }}\,,\quad
\delta c^0_i= \sqrt{2\kappa}\, \bar\rho_i c^0- \frac{\omega_i c_1}{\sqrt{2\kappa -1}}\,, \quad
\delta c_1=-\sqrt{2\kappa -1}\,\bar\omega^k c^{0}_k\,,
\eea
where $\bar\omega^i, \omega_i$ and $\rho^i, \bar\rho_i$ are the parameters associated with
the supercharges $\tilde S^1_i, \tilde{\bar S}^i_1$ and $\tilde  S^2_i, \tilde{\bar S}^i_2\,$.

As the final topic of this section, we indicate which $SU(2|2)$ multiplets the wave functions form for different values of $\ell$.

In general, irreps of $SU(2|2)$ are characterized  by the triple \cite{Bn}
\be
\langle m, n; \vec{C}\rangle\,,
\ee
where the non-negative integers $m, n$ are Dynkin labels of subalgebra $\textit{su}(2)\oplus \textit{su}(2)$ (in the considered
case they are twice the external isospins of the wave superfunctions) and $\vec{C}$ represents three central charges.
The special case is the ``short'' irreps, with
\be
\langle m, n; \vec{C}\rangle\,, \quad \vec{C}^2=\frac{1}{4}\left(m+n+1\right)^2.
\ee
It turns out that our wave superfunction multiplets belong just to this restricted class of the $su(2|2)$ representations.

At the levels $\ell=0$ and $ \ell=2$, the $su(2|2)$ algebra has the central charge $Z=1/2$.
Hence the relevant wave superfunctions encompass the $2|2$ multiplets characterized by the triple
\be
\langle 0, 0; \vec{C}\rangle \,, \quad  \vec{C}^2 = Z^2 = \frac{1}{4}\,.
\ee
This option corresponds to the fundamental representations of $su(2|2)$.

At the level $\ell =1$ the wave superfunction multiplet comprises 4 bosonic and 4 fermionic components,
with  the central charge $Z=1$. It is characterized by the triple
\be
\langle 1, 0; \vec{C}\rangle\,, \quad \vec{C}^2 = Z^2 = 1\,.
\ee
With respect to the bosonic subalgebra $\textit{su}(2)\oplus \textit{su}(2)$, the
bosonic components of the $\ell=1$ supermultiplet are in $(0,0) \oplus (1,0)$, while the fermionic ones in $(1/2, 1/2)\,$.
This means that, with respect to the first $SU(2)$, the bosonic fields are split into the singlet
$c_+$ and the triplet $c^{[2]}_1, c^{[1]}_2, c_-\,$, where
$$
 c_+=\frac{1}{2}\left( c^{[1]}_1+c^{[2]}_2\right),\qquad c_-=\frac{1}{2}\left( c^{[1]}_1-c^{[2]}_2\right).
$$
With respect to the second $SU(2)$, all these components are just singlets. The fermionic components constitute
doublets with respect to both $SU(2)$ groups. To be more precise,
either of $c^{[k]}$ and $c_2^{[k]}$ is a doublet of the first $SU(2)$ acting on the index $[k]$, while another $SU(2)$
combines them both into its doublet (irrespective of the value of $[k]$).

Let us briefly address the case $\kappa < 0$. With the condition $\kappa < -1/2$ taken into account,
the norms \p{Norm1} and \p{Norm2} are positive, while the first level norm \p{Norm3} is negative (we leave aside
the degenerate case $|\kappa| =1/2\,$). The metric operator chosen in the form
\be
G=1+ \frac{8\kappa^2 - 2H^2}{1 -4\kappa^2}\label{-G}
\ee
has the eigenvalues
\be
G=(-1)^{\ell} \lb{-l}
\ee
and so makes all norms positive-definite. The implications of this metric operator radically differ from those of $G$ for $\kappa
> 1/2$. In particular, it commutes with all symmetry generators and so does not affect their Hermitian conjugation
properties. Thus, as opposed to the $\kappa >0$ case, it cannot be directly employed for construction of
the $SU(2|2)$ generators acting in the relevant Hilbert space. Nevertheless, the unitary norms
for $\kappa < 0$ are still invariant under the appropriately defined  $SU(2|2)$ symmetry.
The corresponding generators are obtained by making the substitutions
\be
    \kappa =|\kappa | \,,\quad \ell= n-\ell' \,,\quad S_i^a = \Pi_i^a \,,\quad \bar{S}_a^i = -\bar{\Pi}_a^i
\ee
in the definitions \p{Odin}, \p{Two}. The $su(2|2)$ superalgebra which these generators form is slightly different
from the one defined by the relations \p{alg1} pertinent to the $\kappa > 0$ case. In the basic anticommutator the two sets of
the bosonic $SU(2)$ generators, $J^J_i$ and ${\cal J}^a_b$, switch their places, and in the central charge one
should replace $\kappa \rightarrow |\kappa|\,$, $\ell \rightarrow \ell'$. The 3-vector $\vec{C}$ defined in \p{Cvector}
preserves its form modulo these substitutions.

A different treatment of the $\kappa <0$ case is based on the consideration in the end of subsection 4.1. The
LLL wave superfunction is defined as $\tilde{\Psi}_{\ell' =0} := {\Psi}_{\ell =2}$.
It satisfies the anti-chirality condition
\be
\nabla^{(\kappa)}_j \tilde{\Psi}_{\ell' =0} =  \nabla^{(\kappa)}_j \nabla^{(\kappa-1)}_i\nabla^{(\kappa+1)}_k\Phi^{[ik]}
= 0\quad \rightarrow \quad \tilde{\Psi}_{\ell' =0} =
(1 - \xi\cdot \bar\xi)^{-|\kappa|}\,\bar{\tilde{\Omega}}_0(\bar\xi)\,.
\ee
Passing to the complex-conjugate set of the wave superfunctions, i.e. $\tilde{\Psi}_{\ell'} \rightarrow
\tilde{\Psi}_{\ell'}^\star\,,$ $\tilde{\Psi}_{\ell'=0}^\star=(1 - \xi\cdot \bar\xi)^{-|\kappa|}\,\tilde{\Omega}_0(\xi)\,$,
reduces the $\kappa < 0$ case to the already studied $\kappa > 0$ one.
The relevant $SU(2|1)$ generators are obtained just through the replacement $\kappa \rightarrow |\kappa|\,$ in the $\kappa >0$
expressions, and the ``passive'' form of the supertransformation of the wave superfunctions mimics \p{supop}
\be
\delta^* \tilde{\Psi}^{\star}(\xi, \bar\xi) =
|\kappa|\left(\epsilon \cdot \bar{\xi}+ \bar{\epsilon}\cdot \xi \right)\tilde{\Psi}^{\star}(\xi, \bar\xi)\,.
\ee
Thus the structure of the quantum Hilbert space in the $\kappa < 0$ case is the same as for $\kappa > 0$.
For the ``physical'' values $\kappa < -1/2$, the metric operator $G$ making all norms positive-definite
is of the same form as in \p{Ggen}, up to the substitutions $\kappa \rightarrow |\kappa|$ and $\ell \rightarrow \ell'\,$.
The hidden $SU(2|2)$ symmetry generators leaving invariant the unitary norms
are obtained from \p{Odin}, \p{Two} through the same substitutions.
\setcounter{equation}{0}

\section{The planar limit}
In this section we consider the planar limit of the $SU(n|1)/U(n)$ coset model.

We introduce a scale parameter $r$, rescale the odd variables and the time coordinate in the action corresponding
to the Lagrangian \p{Mflag} as
\be
\xi \rightarrow \xi/r\,, \;\; t \rightarrow t/r^2\,, \;\; \kappa \rightarrow \kappa r^2\,,\lb{Resc}
\ee
and then send $r \rightarrow \infty\,$. The Lagrangian \p{Mflag} goes over to
\be
L=- \dot \xi ^i\dot{\bar{\xi}}_i+i\kappa (\dot \xi^i\bar \xi _i+\dot{\bar\xi } _i\xi ^i)\,. \label{lagr}
\ee
This is equivalent to the following redefinition of the K\"ahler superpotential
\be
K=r^2\log\left(1-\frac{\xi\cdot\bar\xi}{r^2}\right).\label{Kahler-planar}
\ee
In the planar limit $K$ becomes $\xi\cdot\bar\xi\,$, and the target metric and gauge connections read
\be
g^i_j=\partial _j\partial ^{\bar i} K= \delta ^i_j\,,
\ee
\be
A_i=-i\partial _iK= i\bar\xi _i\,, \quad \bar{A}{}^i=i\partial ^{\bar{i}}K= i\xi ^i\,.
\ee

The Hamiltonian is rescaled as
\be
H (\xi, \bar\xi, \kappa)\; \rightarrow \; 1/r^2 H(\xi/r, \bar\xi/r, \kappa r^2) \lb{rescH}
\ee
and in the planar limit becomes
\be
H=\nabla _i^{(\kappa)}\nabla ^{(\kappa)\bar i} - \kappa n := H' - \kappa n\,,\lb{Hamplan}
\ee
where now
\be
\nabla _i^{(\kappa)}=\partial _i+\kappa\bar\xi _i\,,\quad \nabla ^{(\kappa)\bar i}=\partial ^{\bar i}+\kappa\xi ^i\,.
\ee
The covariant derivatives obey the anticommutation relation
\be
\{\nabla _i^{(\kappa)},\,\nabla ^{(\kappa)\bar j}\}=2\kappa\delta ^{j}_i\,.\label{planticom}
\ee

The Lagrangian \p{lagr} enjoys the symmetry under odd magnetic translations with the generators
\be
Q_i=-\partial _i+\kappa\bar\xi _i\,, \quad Q^{\dagger i}=-\partial ^{\bar i}+\kappa\xi ^{i}\,, \quad
\{Q_i, Q^{\dagger k} \} = -2\kappa \,\delta^{k}_i\,,\;\;\{Q_i, Q_{k} \} = 0\,, \lb{Nn}
\ee
as well as a symmetry under $U(n)$ rotations of the coordinates $\xi^i\,$, which define ``$R$-symmetry'' of
the superalgebra \p{Nn}. The full symmetry structure is an obvious contraction
of the superalgebra $su(n|1)\,$. The spinor generators form just an extended ${\cal N}=n\,$, $d{=}1$ Poincar\'e superalgebra,
with the central charge $-2\kappa$ in place of the Hamiltonian appearing in the standard extended supersymmetric mechanics
models (see, e.g., \cite{superc}). The Hamiltonian \p{Hamplan} admits the Sugawara-type representation
\be
H = Q_iQ^{\dagger i} - 2\kappa F + \kappa n\,,
\ee
where $F = \xi^i\partial_i - \bar\xi_i\partial^{\bar i}$ is the $U(1)$ generator. After putting $\xi^{\hat i} =0\,,
\hat i = 1, \ldots n-1\,$,  and replacing $\kappa \rightarrow -\kappa$, for the remaining
fermionic variable $\xi^n$ the Lagrangian \p{lagr} is reduced to that of the ``toy'' fermionic Landau model
of refs. \cite{IMT} and \cite{CIMT}. Also note that \p{lagr}, up to some redefinitions, in the $n{=}2$ case coincides
with the pure fermionic truncation of the
Lagrangian of ${\cal N}{=}4$ super Landau model recently constructed in \cite{N4}.

Let us sketch some salient features of the quantum theory. For simplicity, we will mainly concentrate on the case with $\kappa >0$.

The ground state wave superfunction is defined by the chirality condition $\nabla ^{(\kappa)\bar i}\Psi _0=0\,$,
which amounts to expressing this superfunction through the holomorphic function
\be
\Psi_0=e^{\kappa \xi\cdot\bar\xi }\Omega _0(\xi)\,.\lb{planLLL}
\ee

The higher LL wave superfunctions are obtained as:
\be
\Psi_\ell=\nabla _{i_1}^{(\kappa)}\cdots\nabla _{i_\ell}^{(\kappa)}\Phi ^{[i_1\cdots i_\ell]}, \quad
\Phi ^{[i_1\cdots i_\ell]} = e^{\kappa \xi\cdot\bar\xi }\Omega ^{[i_1\cdots i_\ell]}(\xi)\,,\lb{planLL}
\ee
with the energy of the $\ell\,$th Landau level  being $E_\ell=2\kappa \ell\,$. This formula for $E_\ell$ can be directly
reproduced from \p{spectrum}, making there rescaling \p{rescH} and taking the
planar limit $r \rightarrow \infty\,$. Note that the exponential prefactors in \p{planLLL} and \p{planLL} can
be regarded as the $r\rightarrow \infty$ limit of the prefactor in \p{LLL}, \p{1LL} and \p{chirn}:
$\left(1- r^{-2}\,{\xi}\cdot\bar{\xi}\right)^{-\kappa r^2}\rightarrow e^{\kappa \xi\cdot\bar\xi }\,$.

The invariant norm is defined by
\be
||\Psi ||^2=\int d\mu _0\Psi ^{\star } \Psi\,,\label{fnorms}
\ee
where $\int d\mu _0 =\prod \partial _i\partial ^{\bar i}$.
Using the anticommutation relation \p{planticom}, it is straightforward
to compute the norm of the $\ell\,$th level wave superfunction:
\be
||\Psi _\ell ||^2=\ell!(2\kappa)^\ell\int d\mu_0\Phi_{[i_1\cdots i_\ell]}^{\star }\Phi^{[i_1\cdots i_\ell]}
=\ell!(2\kappa)^\ell\int d\mu_0 e^{2\kappa {\xi}\cdot\bar{\xi}}\Omega_{[i_1\cdots i_\ell]}^{\star}\Omega^{[i_1\cdots i_\ell]}\,.
\ee
This integral is obviously not positive-definite, for both $\kappa >0$ and $\kappa < 0\,$.
To make all norms positive, we should redefine the inner product:
\be
\langle\langle\Psi |\Phi\rangle\rangle =\int d\mu\Psi ^\star G\Phi\, .
\ee
The metric operator $G$ is defined as
\bea
&&1)\quad G=\left(1-2\hat{m}_1\right)\ldots\left(1-2\hat{m}_n\right)\,,\qquad \text{for}\quad \kappa > 0\,,\label{planG1} \\
&&2)\quad G=\left(1-2\hat{n}_1\right)\ldots\left(1-2\hat{n}_n\right)\,,\qquad \;\;\text{for}\quad \,\kappa < 0 \,.\label{planG}
\eea
Here
\bea
&&\hat{m}_i= \frac{\nabla _i^{(\kappa)}\nabla ^{(\kappa)\bar i}}{2\kappa}+\left(\xi^i\partial_i
-\bar {\xi }_i \partial^{\bar {i}}\right), \nn
&&\hat{n}_i=-\frac{\nabla _i^{(\kappa)}\nabla ^{(\kappa)\bar i}}{2|\kappa|}\,,\label{ni}
\eea
and no summation over the index $i$ is assumed.

The operator $G$ anticommutes with the supercharges for $\kappa>0$,
\be
    \lbrace Q_i, G\rbrace =0\,,\qquad \lbrace Q^{\dagger i}, G\rbrace =0\,,
\ee
and commutes with them for $\kappa<0\,$. Note that, within
our conventions, just the latter option corresponds to the ``toy'' $SU(1|1)$ invariant
fermionic Landau model considered in refs. \cite{IMT,CIMT}. The
relevant metric operator is the $n=1$ case of \p{planG}. The
alternative choice of $\kappa$, with the metric operator being the
$n=1$ case of \p{planG1}, was not addressed in these papers.

Though in the notation \p{planG1}, \p{planG} covariance with respect
to the $U(n)$ $R$-symmetry is not manifest, one can check that the
operators $G$ commute with $U(n)$ generators. For instance, in the $n{=}2$ case and for $\kappa <0$
the metric operator $G$ can be rewritten in the manifestly
$U(2)$ covariant form as
$
G=1+ \frac{1}{2}|\kappa|^{-2}{H'}^2 + 2 |\kappa|^{-1} {H'}\,.
$
This operator can be obtained as the planar $r\rightarrow \infty $ limit of the metric operator \p{-G}, in which
$\xi_i$ and $\kappa$ are rescaled according to \p{Resc}.
Upon truncation to $n=1\,$, we find that ${H'}^2 \rightarrow
-2|\kappa| {H'}$ and $G \rightarrow 1 + |\kappa|^{-1}{H'}\,.$
This $G$ is the metric operator of the fermionic model of refs. \cite{IMT,CIMT}.

Finally, let us illustrate the general consideration by the example of the positive norm for the
LLL wave function in the $n{=}2$ case with $\kappa>0$:
\be
     ||\Psi _0||^2=\bar c_1 c_1+2\kappa\bar c^{0i}c^0_i+(2\kappa)^2\bar c^0 c^0\,.
\ee
The coefficients in the $\xi^i$ expansion of the holomorphic LLL superfunction $\Omega_0(\xi)$ were defined
precisely as in \p{defc0c1}.

\subsection{Symmetries of the quantum planar $n{=}2$ model}
The superplane Landau models obtained as a large radius limit of the $SU(2|1)/U(1|1)$ supersphere or the $SU(2|1)/[U(1)\times U(1)]$
superflag Landau models respect the worldline ${\cal N}{=}2$ supersymmetry which emerges as a contraction of the appropriate
extensions of $SU(2|1)$ realized in the Hilbert space of the quantum curved models \cite{BCIMT}.

In the planar limit, no any worldline supersymmetry appears in the $n{=}2$ odd-coset model. To see what kind of the
algebraic structure is recovered in this limit from the $su(2|2)$ generators defined in subsection 6.2, we should make,
in all odd generators \p{Odin},
\p{Two}, the rescaling of $\xi, \bar\xi, \kappa$ according to \p{Resc} and multiply them by the additional factor $1/r$ ,
while no such a factor is needed in the case of the bosonic generators \p{BosOne}, \p{BosTwo} (it is enough to make the rescaling
of the odd variables and $\kappa$ in them). Keeping in mind that, to the leading order,  $\sqrt{C_2} = 2\kappa r^2 + \ldots $
for any level $\ell$, it is easy to check that in the limit $r \rightarrow \infty$ all odd $SU(2|2)$ generators become
either $\partial_i$ or
$2\kappa \xi^i$ (modulo signs), which are just the generators $Q_i$ or $Q^{\dagger i}$
in the holomorphic representation, with the only non-vanishing anticommutator $\{Q, Q^\dagger\}$, as is given in \p{Nn} (for $n{=}2$).
Thus no new fermionic generators appear in this limit. The bosonic generators \p{BosOne} retain their form and define the $SU(2)$
$R$-symmetry group of the flat $n{=}2$ superalgebra \p{Nn}. The extra $SU(2)$ generators \p{BosTwo} also retain their form, modulo
the substitution $\sqrt{C_2} \rightarrow 2\kappa\,$. They generate the second $R$-symmetry $SU(2)$ group
of the flat $n{=}2$ superalgebra, such that the generators $Q_i$ and $Q^{\dagger i}$ are mixed under this $SU(2)$.

Surprisingly, it is still possible, at least for the choice of $\kappa > 0$, to show that the space of quantum states
of the $n{=}2$ fermionic planar Landau model exhibits $SU(2|2)$ symmetry.

For $\kappa >0\,$, the Hermitian conjugates of the supercharges $Q_i$ are given by
\be
Q^{\ddagger i} \equiv G Q^{\dagger i}G =- Q^{\dagger i}\,,\qquad i=1,2\,.
\ee
The metric operator can be used to construct another pair of the supercharges
\bea
&&\left(Q_G\right)_i  \equiv \frac{1}{2}\left[Q_i, G\right]=Q_i G = -GQ_i\,.
\eea
These two pairs can be combined into the complex quartet supercharges $\hat{S}_i^a$ as
\bea
    \hat{S}_1^1 &=& \frac{1}{2}\left(2\kappa\right)^{-\frac{1}{2}}\left(1+G\right) Q_{1}\,,\qquad
    \hat{S}_1^2 = -\frac{1}{2}\left(2\kappa\right)^{-\frac{1}{2}}\left(1+G\right) Q^{\ddagger 2}\,,\nn
    \hat{S}_2^1 &=& \frac{1}{2}\left(2\kappa\right)^{-\frac{1}{2}}\left(1+G\right) Q_{2}\,,\qquad
    \hat{S}_2^2 = \frac{1}{2}\left(2\kappa\right)^{-\frac{1}{2}}\left(1+G\right) Q^{\ddagger 1}\,.\label{plansu}
\eea

These supercharges together with their conjugates $\hat{S}{}^{\ddagger i}_a$ satisfy just the (anti)commutation relations  \p{alg2}
of the superalgebra $su(2|2)\,$, with $Z= 1/2\,$ and the $SU(2)$ generators $(\hat{J}_\pm, \hat{J}_3)\,$,
$(\hat{\cal J}_\pm\,, \hat{\cal J}_3)$ defined as
\bea
&&\hat{J}_+ =\left(2\kappa\right)^{-1} Q^{\ddagger 1}Q_2,\quad \hat{J}_- =\left(2\kappa\right)^{-1} Q^{\ddagger 2}Q_1, \quad
\hat{J}_3 =\frac{1}{2}\left( \hat{m}_1 - \hat{m}_2\right)\,,\nn
&&\hat{\cal J}_+=\left(2\kappa\right)^{-1} Q^{\ddagger 1}Q^{\ddagger 2}\,,\quad \hat{\cal J}_-=\left(2\kappa\right)^{-1} Q_2 Q_1\,,
    \quad \hat{\cal J}_3= -\frac{1}{2} \left(1- \hat{m}_1 - \hat{m}_2\right).\label{planB}
\eea

It is instructive to rewrite the first set of the $SU(2)$ generators in the covariant form
\bea
\hat{J}_j^i=\frac{1}{4\kappa}\left(\left[Q^{\ddagger i}, Q_j\right]-\frac{1}{2}\delta_j^i
\left[Q^{\ddagger k}, Q_k\right] \right),
\qquad \hat{J}_+= \hat{J}_2^1\,,\; \hat{J}_-= \hat{J}_1^2\,,\; \hat{J}_3=\hat{J}_1^1=-\hat{J}_2^2\,.\lb{covI}
\eea
We see that they do not coincide with the generators of the $R$-symmetry $SU(2)$: in the holomorphic basis, with $Q_i = -\partial_i,
Q^{\ddagger i} = -2\kappa \xi^i$, one has
\be
\hat{J}_j^i = \xi^i\partial_j - \frac{1}{2}\delta^i_j\,\xi^k\partial_k\,, \lb{holomhatJ}
\ee
which should be compared with the $R$-symmetry $SU(2)$ generators \p{BosOne} in the same basis (their form is preserved
upon passing to the planar limit). The difference is the absence of the matrix parts $\hat B$ in \p{holomhatJ}
for any level $\ell$ as compared to the generators \p{BosOne} which necessarily include the $\hat B$ parts for $\ell =1$. Nevertheless, the
positive norms are invariant under both these $SU(2)$ symmetries separately. The $R$-symmetry $SU(2)$ can be interpreted as one of the
outer automorphisms of the considered $su(2|2)$ superalgebra: it uniformly rotates the doublet indices $i, j, k$ of
the fermionic generators $\hat{S}^a_i$ and the bosonic generators $\hat{J}^i_j\,$.

As for the second set of $SU(2)$ generators in \p{planB}, they coincide with those of the second $R$-symmetry
group $SU(2)$ and have, in the holomorphic basis, the same form as in \p{BosTwo}, up to the substitution
$\sqrt{C_2} \rightarrow 2\kappa\,$.

Note that the pair of generators $(Q_G)_i = Q_iG\,, \;(Q_G)^{\ddagger i} = - Q^{\ddagger i}G$ form the
same flat algebra $\{ Q_G, Q_G^\ddagger\} = 2\kappa$ as the pair $Q_i, Q^{\ddagger i}\,$. Their crossing anticommutators
produce the $su(2)$ generators \p{planB}. The fermionic generators $Q, Q_G$ and their conjugates
just fix another basis in the odd sector \p{plansu} of  the $su(2|2)$ superalgebra (this basis was employed, e.g., in \cite{N4}).

The LLL and the second LL wave superfunctions belong to the short representation of this $su(2|2)$
\be
    \langle 0, 0; \vec{C}\rangle\,,
\ee
with $\vec{C}=(1/2, 0, 0)$. The first level is described by the direct sum of such representations
\be
    \langle 0, 0; \vec{C}\rangle\oplus \langle 0, 0; \vec{C}\rangle\,.
\ee

\section{Summary and outlook}

In this paper, we continued the study of super Landau models associated with the supergroup $SU(n|1)\,$. This study was initiated
in \cite{Odd,CPn1} for an arbitrary $n$ and further performed in more detail for $n{=}2$ in \cite{IMT2,BCIMT}. We  constructed
the model on the pure odd coset $SU(n|1)/U(n)$ as a generalization of the consideration of \cite{CPn1}
which dealt only with the lowest Landau
level sector of such a model. An important peculiarity of this model is the finite number $n+1$ of Landau levels in the sector
spanned by the wave superfunctions with the vanishing external $SU(n)$ ``spin''. We presented the action of the model,
as well as its quantum Hamiltonian, for an arbitrary $n$,  found the energy spectrum  and defined the relevant wave superfunctions.
For the particular case of $n{=}2$ we showed that the space of quantum states of the model reveals hidden $SU(2|2)$ symmetry:
at each Landau level $\ell = 0, 1, 2$, the relevant wave superfunctions constitute ``short'' $SU(2|2)$ multiplets.
Like in other super Landau models, the Hilbert space for any $n$ includes the states the norms of which are not positive-definite;
for $n{=}2$ we gave the explicit form of the ``metric'' operator which modifies the inner product in such a way
that the norms of all states become non-negative, thus demonstrating that the model is unitary. Such operator is expected
to exist for any $n$. We also studied the planar limit of the   $SU(n|1)/U(n)$ models. In this limit,
the superalgebra $su(n|1)$ converts into a superalgebra which is isomorphic to
$n$-extended $d{=}1$ Poincar\'e superalgebra, with a central charge playing the role of the $d{=}1$ Hamiltonian. We presented,
for an arbitrary $n$, the explicit form of the metric operator ensuring the norms of all quantum states to be non-negative.
For $n{=}2$, we find out that the Hilbert space for all three levels carries a dynamical hidden $SU(2|2)$ symmetry,
like in the  $SU(2|1)$ model, though these two $SU(2|2)$ symmetries are realized in entirely different ways.

Our consideration shows that the appearance of the hidden $SU(2|2)$ symmetries in the superflag and supersphere Landau
models based on the supercosets $SU(2|1)/[U(1)\times U(1)]$ and $SU(2|1)/U(1|1)$ \cite{BCIMT} is not accidental:
the same phenomenon persists in the Landau model on the pure
odd supercoset $SU(2|1)/U(2)$. Moreover, the planar limit of this model also surprisingly exhibits $SU(2|2)$ symmetry
on the quantum states (as a substitute of the worldline ${\cal N}{=}2$ supersymmetry of the planar limits
of the superflag and supersphere models \cite{CIMT,SupLan}). It would be interesting to inquire whether
the Hilbert spaces of the odd-coset Landau  models with $n{>}2$ and their planar limits admit any extended
hidden symmetry.

The presence of hidden $SU(2|2)$ symmetry in the quantum $SU(2|1)$ Landau models (or some other symmetries of the similar kind
in the case of Landau models based on further extensions of $SU(2|1)$) suggests the possible relation of this class
of models to such integrable systems as the $su(2|2)$ or $su(3|2)$ spin chain models  and, finally,
to ${\cal N}{=}4, d{=}4$ super Yang-Mills theory and string theory (see, e.g., \cite{Bn,AFr,schain,mih}).
It would be also interesting to retrieve the super Landau models via dimensional reduction from
some higher-dimensional sigma models with the supergroup target spaces.

Regarding possible physical applications of the odd coset Landau models, we would like to express a hope
that the latter (or their analogs associated with other supergroups) could be relevant to the description
of the quantum Hall effect and its spin extensions \cite{Hasebe2,Hasebe,QHE}. In a sense,
the quantum fermionic Landau model could be regarded as a sort of the ``parent model'' for those associated with other
$SU(n|1)$ supercosets.  This is based on the following reasonings. The maximal linearly realized symmetry
of the $SU(n|1)/U(n)$ model is $U(n)$, and we could
couple the fermionic coset variables to some $U(n)$ ``matter'' multiplets with preserving the full nonlinear $SU(n|1)$
symmetry, following the general recipes of the nonlinear realizations theory. With adding the proper $U(n)$
(and $SU(n|1)$) invariant potentials to the action, we could hope to trigger the spontaneous breaking of $U(n)$ down to
a smaller symmetry $H\subset U(n)$, so that the original fermionic variables, together with the bosonic $U(n)/H$ ones, parametrize
some ``mixed'' fermionic-bosonic supercosets $SU(n|1)/H$. For instance, the superflag $SU(2|1)/[U(1)\times U(1)]$ model
could be recovered through the spontaneous breaking pattern $U(2) \rightarrow H=U(1)\times U(1)\,$.
The presence of additional bosonic ``sigma fields'' in such extended models as compared to the pure coset $SU(n|1)/H$ case,
with non-trivial potential terms, could drastically change the quantum properties of these models. The relevant
$d{=}1$ supercoset models should follow from the extended models in the ``long-wave'' limit, in the same way
as the higher-dimensional nonlinear sigma models follow from the linear ones.

\setcounter{equation}{0}

\section*{Acknowledgements}
E.I.  thanks Sergey Fedoruk and Askold Perelomov for discussions and useful correspondence.
We acknowledge a partial support from the RFBR grants Nr. 12-02-00517 and Nr. 11-02-90445.

\setcounter{equation}{0}
\appendix
\section{One-dimensional sigma models from gauging}

In this Appendix we demonstrate how the invariant $d{=}1$ actions of some other Landau-type models
can be recovered by the gauge method of Section 2.\\

\subsection{Landau model on the supersphere $SU(2|1)/U(1|1)$}

Superspherical Landau model describes a motion of a charged particle on the supersphere $SU(2|1)/U(1|1)$ parametrized
by one complex even and one complex odd coordinates \cite{BCIMT}.

Consider the set of two complex bosonic fields $u^i, i=1,2$, and one complex fermionic field $\xi$ which form a
fundamental representation of $SU(2|1)$. The $SU(2|1)$ transformations are defined as a group of linear transformations
leaving invariant the following bilinear form:
\be
u^1\bar {u}_1 +u^2\bar {u}_2 +\xi \bar {\xi }=inv\,.
\ee

As the first step, we impose the manifestly $SU(2|1)$ covariant constraint
\be
u^1\bar {u}_1 +u^2\bar {u}_2 +\xi \bar {\xi }=1\,,
\ee
from which we eliminate $|u^1|$ as
\be
|u^1|=\sqrt {1-u^2\bar {u}_2 -\xi \bar {\xi }}\,.
\ee
As the second step, we
gauge the $U(1)$ symmetry which acts as a multiplication of the fields $(u^i, \xi)$ by the common phase and
commutes with $SU(2|1)\,$.
The sigma-model type action, which is invariant under both $SU(2|1)$ and gauge $U(1)$ symmetries, is given by
\be
L=\nabla u^i\bar {\nabla }\bar {u}_i +\nabla \xi \bar {\nabla }\bar {\xi}+2\kappa {\cal A}\, , \label{L1}
\ee
where we have introduced the covariant derivatives $\nabla =\partial _t -i {\cal A}$,  $\bar {\nabla }=\partial _t +i {\cal A}$
involving
the non-propagating $U(1)$ gauge field, ${\cal A}(t)$, and added the Fayet-Iliopoulos term for the gauge field $\sim \kappa\,$.

The local $U(1)$ invariance allows us to gauge away a phase of $u^1$ and thus to make $u^1$  real.
In this gauge
\be
u^1= |u^1| = \sqrt {1-u^2\bar {u}_2 -\xi \bar {\xi }}\,, \nn
\ee
and we are left with the complex bosonic and fermionic fields $u^2$ and $\xi$ as the only independent degrees of freedom.
The field ${\cal A}(t)$ enters \p{L1} without derivatives, so we can eliminate it by its algebraic equation of motion:
\be
{\cal A}(t) =\frac{i}{2}(u\cdot\dot{\bar u}-\dot{u}\cdot\bar u+\xi\dot{\bar\xi}-\dot{\xi}\bar\xi-2\kappa)\,.\label{sphA}
\ee
It is also convenient to pass to the new independent variables, Grassmann-even $z(t)$ and Grassmann-odd $\zeta (t)$,
\bea
(u^2 , \xi )\to (z, \zeta )\,, \quad u^2 =\frac{z}{\sqrt {1+z\bar {z}+\zeta \bar {\zeta }} }\,, \;\;
\xi=\frac{\zeta}{\sqrt{1+z\bar {z}+\zeta \bar {\zeta }}}\,.
\label{ssph u^1}
\eea
Then the Lagrangian \p{L1}, with $A(t)$ being expressed in terms of $z, \zeta$ by \p{sphA} and \p{ssph u^1}, can be written as
\be
L=g_{\bar {B}A}\dot {Z}^A\dot {\bar {Z}}^B+\kappa \left( {\dot {Z}^AA_A +\dot {\bar {Z}}^BA_{\bar {B}} } \right),\label{L2-sph}
\ee
where gauge connections are $ A_A =-i\partial _A K\,,\;  A_{\bar {A}} =i\partial_{\bar {A}} K\,, \;
K=\ln (1+z\bar {z}+\zeta \bar {\zeta })\,$, and
the metric on supersphere is
\bea
&&g_{\bar {z}z} =\frac{1+\zeta \bar {\zeta }}{\left( {1+z\bar {z}+\zeta \bar {\zeta }} \right)^2}\,,
\quad\quad g_{\bar {z}\zeta } =-\frac{z\bar {\zeta }}{\left({1+z\bar {z}} \right)^2}\,,\\
&&g_{\bar {\zeta }z} =\frac{\zeta \bar{z}}{\left( {1+z\bar {z}} \right)^2}\,,\quad\quad g_{\bar {\zeta }\zeta }
=\frac{1}{1+z\bar {z}}\,.
\eea

The metric can be concisely written as $g_{\bar {B}A} =\partial _{\bar {B}} \partial _A K\,$,
i.e. the function $K(z, \bar z, \zeta, \bar\zeta)$
is the corresponding super K\"ahler potential. The Lagrangian \p{L2-sph} coincides with the one found in \cite{BCIMT}
by a different method. It should be
pointed out that the WZ term in \p{L2-sph} originates from the FI term in the original gauge action \p{L1}, like in the
odd coset sigma model Lagrangian \p{Mflag}, \p{odd-L1}.

\subsection{Landau model on $SU(n)/U(n-1)$}
Our second example is the purely bosonic extended Landau-type model on
the coset space $SU(n)/U(n-1)$, which is a generalization of the $S^2 \sim \mathbb{CP}^1$ Haldane model \cite{Haldane}.

Consider bosonic multiplet $u^\alpha (t),\;\alpha =1,...,n$ in the fundamental representation of the $SU(n)$.
This group acts on these $d{=}1$ fields as
\be
\delta u^\alpha = \lambda^\alpha_\beta u^\beta, \quad \overline{(\lambda^\alpha_\beta)} = -\lambda^\beta_\alpha\,, \;
\lambda^\alpha_\alpha = 0\,.
\ee

Impose the $SU(n)$ invariant constraint
\be
\bar u_{\alpha} u^\alpha = 1 \label{invc}
\ee
and define the $SU(n)$ invariant Lagrangian
\be
L=\bar {\nabla }\bar {u}_\alpha \nabla u^\alpha +2\kappa {\cal A}\,.  \label{inlagr}
\ee
Here the auxiliary gauge field ${\cal A}(t)$ ensures the local $U(1)$ invariance of \p{inlagr},
the corresponding gauge-covariant derivatives being defined as $\nabla =\partial_t -i{\cal A}$ and $\bar {\nabla }
=\partial_t +i{\cal A}\,$.
As in other examples, we may eliminate ${\cal A} (t)$  by its algebraic equation of motion,
\be
{\cal A} = \frac{i}{2}\left(\dot{\bar{u}}{}_\alpha u^\alpha - \bar{u}_\alpha\dot{u}{}^\alpha + 2i\kappa\right).\lb{Alast}
\ee
We can also make use of the local $U(1)$ invariance to choose the $u^1$ field real.
Then, with making use of the constraint \p{invc}, this field can be expressed through the remaining ones as
$$
u^1 =\sqrt {1-\bar {u}_a u ^a }\,.
$$

The natural realization of the $SU(n)$ group as a group of left shifts on the coset space
$SU(n)/U(n-1)$ is in terms of the complex coordinates $z^a,\;a=1,...,n-1\,,$
with the holomorphic $SU(n)$ transformations:
\be
\delta z^a =\lambda ^a_1 +\lambda ^a_b z^b -\lambda ^1_1 z^a -\lambda ^1_b z^a z^b\,.   \label{z^a tr}
\ee
Coordinates with such a transformation law correspond to the realization of the coset space $SU(n)/U(n-1)$
as the complex projective space $\mathbb{CP}^n\,$. The connection between the new coordinates $z^a$ and the old
coordinates $u^\alpha$ is as follows
\be
u^a =\frac{z^a }{\sqrt {1+z^a \bar {z}_a } }\,,
\ee
and therefore
\be
u^1=\frac{1}{\sqrt {1+z^a \bar {z}_a}}\,.
\ee

In terms of the new coordinates, after substituting the expression \p{Alast} for ${\cal A}(t)$ back to the Lagrangian \p{inlagr},
the latter takes the form
\be
L=\frac{\dot {z}^a \dot {\bar {z}}_a }{1+z^b \bar {z}_b }-\frac{\dot {z}^a\bar {z}_a z^b \dot {\bar {z}}_b }{(1+z^c \bar {z}_c)^2}-
i\kappa\frac{\dot {z}^a \bar {z}_a -z^a \dot {\bar {z}}_a }{1+z^b\bar {z}_b }\,. \lb{lagr2}
\ee
In the $n{=}2$ case it is reduced to the Lagrangian of Haldane model, whereas for $n=3$ it is the Lagrangian used in \cite{Kara}
for description of a variant of the four-dimensional quantum Hall effect. Note that the coefficient in front of the $U(1)$ WZ term
in \p{lagr2} comes from the FI term in \p{inlagr}, like in other examples.

\end{document}